\documentclass{article}

\usepackage{PRIMEarxiv}

\usepackage[utf8]{inputenc} 
\usepackage[T1]{fontenc}    
\usepackage{hyperref}       
\usepackage{url}            
\usepackage{booktabs}       
\usepackage{amsfonts}       
\usepackage{nicefrac}       
\usepackage{microtype}      
\usepackage{lipsum}
\usepackage{fancyhdr}       
\usepackage{graphicx}       
\graphicspath{{media/}}     

\usepackage{algorithmic}
\usepackage{textcomp}
\usepackage{xspace}
\usepackage{multirow}
\usepackage{wrapfig}
\usepackage{float} 
\usepackage{tikz}
\usepackage{paralist}
\usepackage{subcaption}
\usepackage{amsmath}
\usepackage{bm}
\usepackage{comment}

\usepackage[most]{tcolorbox}
\usepackage{enumitem}
\usepackage{xcolor}
\usepackage{fontawesome5}

\usepackage{longtable}
\usepackage{colortbl}

\usepackage{tikz}

\usepackage{pifont}

\newcommand{\revision}[1]{{\color{black}#1}}

\newcommand{\circlednum}[1]{%
  \tikz[baseline=(char.base)]{
    \node[shape=circle,draw,inner sep=0.5pt, line width=0.3pt, scale=1] (char) {\sffamily\bfseries\footnotesize #1};
  }%
}
\usepackage[numbers]{natbib}

\usepackage[xindy,acronym]{glossaries}
\glsdisablehyper

\newacronym{fm}{FM}{Foundation Model}
\newacronym{cps}{CPS}{Cyber-Physical System}
\newacronym{llm}{LLM}{Large Language Model}
\newacronym{dt}{DT}{Digital Twin}
\newacronym{av}{AV}{autonomous vehicle}
\newacronym{re}{RE}{requirements engineering}
\newacronym{vlm}{VLM}{Vision-Language Model}
\newacronym{apr}{APR}{Automated Program Repair}

\title{Foundation Models for Software Engineering of Cyber-Physical Systems: the Road Ahead}

\author{
  Chengjie Lu \\
  Simula Research Laboratory and \\ University of Oslo \\
  Oslo, Norway \\
  \texttt{chengjielu@simula.no} \\
   \And
  Pablo Valle \\
  Mondragon University \\
  Mondragon, Spain \\
  \texttt{pvalle@mondragon.edu} \\
   \And
  Jiahui Wu \\
  Simula Research Laboratory and \\ University of Oslo \\
  Oslo, Norway \\
  \texttt{jiahui@simula.no} \\
  \And
  Erblin Isaku \\
  Simula Research Laboratory and \\ University of Oslo \\
  Oslo, Norway \\
  \texttt{erblin@simula.no} \\
  \And
  Hassan Sartaj \\
  Simula Research Laboratory \\
  Oslo, Norway \\
  \texttt{hassan@simula.no} \\
  \And
  Aitor Arrieta \\
  Mondragon University \\
  Mondragon, Spain \\
  \texttt{aarrieta@mondragon.edu} \\
   \And
  Shaukat Ali \\
  Simula Research Laboratory \\
  Oslo, Norway \\
  \texttt{shaukat@simula.no} \\
}

\begin{document}
\maketitle

\begin{abstract}
\Glspl{fm}, particularly \glspl{llm}, are increasingly used to support various software engineering activities (e.g., coding and testing). Their adoption in software engineering of \glspl{cps} is also growing. However, research in this area remains limited. Most existing studies have primarily focused on \glspl{llm}, only one type of \gls{fm}, leaving ample opportunities to explore other \Glspl{fm}, such as vision-language models. We argue that, in addition to \glspl{llm}, other \glspl{fm} utilizing different data modalities (e.g., images, audio) and multimodal models (which integrate multiple modalities) hold great potential for supporting \gls{cps} software engineering, given that these systems process diverse data types. \revision{To address this, in this first systematic effort, we present a forward-looking research roadmap for integrating \glspl{fm} into commonly known phases of \gls{cps} software engineering, thereby making it accessible to most software engineers. We derive the roadmap from the literature, emerging trends in \Glspl{fm}, and gaps identified from the literature. The roadmap highlights key challenges and actionable research opportunities for the software engineering community to guide future research.} \revision{Moreover, we discuss the common challenges associated with applying \glspl{fm} across six dimensions (e.g., technical, economic and resource, and human aspects).}
\revision{This roadmap aims to provide a visionary guide for researchers and practitioners, outlining directions for future work and performing future empirical studies.}
\end{abstract}


\maketitle

\glsresetall

\section{Introduction}\label{introduction}

\Glspl{fm} such as \glspl{llm} and \glspl{vlm} have seen a sharp rise in their use across all walks of life. The use of \Glspl{fm} is expected to grow further in the coming years~\cite{Schneider2024}. According to Precedence Research~\cite{precedence2025llm}, the global language model market size is calculated at USD 7.77 billion in 2025. \revision{Furthermore, the market size is also expected to exceed USD 123.09 billion by 2034~\cite{precedence2025llm}. This market growth is primarily driven by the increasing adoption of \glspl{fm} across diverse industries (e.g., mobility, healthcare, and manufacturing) and by the growing demand for automation and data-driven autonomous decision-making.}

\revision{There is also a significant rise in interest in using \glspl{fm}, particularly \glspl{llm}, for various software engineering problems~\cite{LLMs4SESurvey} across various phases of software engineering. Among these phases, software development, maintenance, and quality assurance phases remain the most dominant areas of research~\cite{LLMs4SESurvey}.} Besides, \glspl{fm} have demonstrated great capabilities in supporting the development of \gls{cps}. For instance, \gls{llm} enables robots (one type of  \gls{cps}) to understand and generate human-like behaviors, plan and perform tasks, and interact more naturally with humans. According to market estimations~\cite {market2025llmrobotics}, the \gls{llm} market in robotics, including applications in industrial robots, service robotics, and humanoid robots, is expected to increase from USD 2.8 billion in 2024 to approximately USD 74.3 billion by 2034.
However, the use of \glspl{llm} and \glspl{fm}, in general, remains limited for the software engineering of \glspl{cps}. Such systems are typically multidisciplinary software systems integrating aspects such as hardware, software, communication, and interaction with the physical environment, including humans, \revision{with applications in various domains such as healthcare, transportation, logistics, and manufacturing.} Therefore, they pose unique challenges for software engineering across all phases of development. 

In the literature, some works identified research challenges and opportunities for using \glspl{llm} in software engineering. Examples include surveys on using \glspl{llm} for software engineering tasks~\cite{LLMs4SESurvey,LLM4SESurvey2} and a survey on \gls{llm}-based agents for software engineering~\cite{LLM-basedAgentforSE}. Furthermore, Hassan et al.~\cite{ReThinkingSE} present a proposal for rethinking software engineering in the era of \glspl{fm}. Instead, we focus on software engineering for \glspl{cps}. Xu et al.~\cite{LLMs-enabledCPS}, introduce a set of research opportunities for \glspl{cps} that use \glspl{llm}. Besides, vision \glspl{fm} such as \glspl{vlm} provide opportunities to support the \gls{cps} development~\cite{cui2024survey} given their ability to handle multimodal data. In contrast, we focus on the software engineering of \glspl{cps} with \glspl{fm}. A related paper~\cite{FMDTIsola}, presents research opportunities for creating \glspl{dt} of \glspl{cps} with \glspl{fm}. \revision{In contrast, we provide a broader, forward-looking roadmap that covers all software engineering phases of \glspl{cps}, integrating \glspl{llm} and other \Glspl{fm} as well as cross-cutting issues, positioning our paper as the first systematic effort to offer a high-level, visionary guide for integrating \Glspl{fm} into \gls{cps} software engineering.

Given that \glspl{cps}, depending on the domain, process a wide range of data, such as images, videos, audio, and their combinations, it is natural to use \glspl{fm} with different data modalities to support their software engineering. As a result, many research opportunities exist in \gls{cps} software engineering, particularly in addressing the challenges associated with using \glspl{fm} and integrating them into \gls{cps} software engineering workflows. \revision{To guide future research in this area, we present a forward-looking research roadmap based on the current state of the art, emerging \glspl{fm} trends, gaps identified in the literature, and recent technological developments.}
\revision{Furthermore, to provide actionable research opportunities and methodological suggestions, we first identify research challenges from the literature and, for each challenge, define corresponding research opportunities framed as key research questions. We then complement each research opportunity with concrete methodological guidance by outlining suitable methodologies, proposing baseline approaches for comparison, and suggesting relevant evaluation metrics to enable rigorous empirical evaluations in the future. We organize challenges and opportunities around typical software engineering phases, including requirements engineering, modeling, and testing, to make them accessible to software engineers with diverse expertise. In addition, we address cross-cutting issues across the phases, such as sustainability and human factors, to provide a holistic view of research opportunities.

Given that the roadmap is intended as a visionary, forward-looking guide, a type of contribution absent from the current literature, we do not follow the procedures of systematic literature reviews. However, where possible, we support our arguments with references to the existing literature to strengthen the roadmap's foundation. Since we didn't follow a systematic literature review process, some challenges and research opportunities may not have been captured. However, as a research roadmap, its primary aim is to present representative challenges and opportunities, guide researchers, and serve as a foundation for future empirical work. Missing some items does not reduce its value, as the roadmap can be iteratively refined as new studies emerge.
}

\textbf{Remark.} Note that this paper focuses on the use of \glspl{fm} to support the software engineering of \glspl{cps}, including requirements engineering, design and modeling, software development, testing, debugging and repair, and evolution. The use of \glspl{fm} embedded within \gls{cps} operation is outside the scope of this work.
}

We organize our paper as follows: After introducing a generic background in Section~\ref{sec:backgroun}, \revision{we present the methodology for developing this roadmap in Section~\ref{sec:method}.}
We then discuss challenges and research opportunities related to requirements engineering for \glspl{cps} software in Section~\ref{sec:RE4CPS}, system design and modeling in Section~\ref{sec:Model4CPS}, software development in Section~\ref{sec:SoftwareDev4CPS}, testing in Section~\ref{sec:CPSTest}, debugging and repair in Section~\ref{sec:CPSrepasir}, and evolution in Section~\ref{sec:CPSevolution}. In each section, we first outline the key challenges encountered when designing CPS software and then highlight opportunities where \glspl{fm} can help address these challenges. Additionally, common challenges related to the application of \glspl{fm} across all \gls{cps} software engineering phases are discussed in Section~\ref{sec:challenges}.


\section{Background}
\label{sec:backgroun}

\subsection{Foundation Models}
\glspl{fm} are large general models pretrained on a massive amount of data and have demonstrated extraordinary performance in a wide range of tasks, including computer vision~\cite{wang2023visionllm,awais2025foundation}, software engineering~\cite{10.1109/TSE.2024.3368208,LLMs4SESurvey}, and robotics~\cite{firoozi2025foundation,zitkovich2023rt}. Different from traditional machine learning models that are trained on dedicated datasets for a specific task, FMs are trained on heterogeneous datasets and can be adapted to a variety of downstream tasks through prompting, fine-tuning, or other adaptation techniques. The emergence of FMs can be traced back to the development of pretrained models in natural language processing~\cite{FMOppor}, where transfer learning makes \glspl{fm} possible by taking knowledge learned from one task to another. Moreover, advances in computing resources, optimization algorithms, and the availability of large datasets have facilitated the construction of \glspl{fm}. 

Transformer~\cite{vaswani2017attention} is a breakthrough architecture that captures long-term dependencies and enables scalable training on massive datasets. It has become the dominant backbone of modern \glspl{fm}, powering models such as BERT~\cite{devlin2019bert}, GPT~\cite{radford2018improving}, and their multimodal variances (e.g., GPT4~\cite{achiam2023gpt}). Typically, the transformer architecture can be used in three different designs, including encoder-only, decoder-only, and encoder–decoder (or sequence-to-sequence) architectures. The encoder-only design, exemplified by BERT~\cite{devlin2019bert}, focuses on creating rich contextual representations of input sequences and is commonly used for understanding tasks such as classification or information retrieval. The decoder-only design, used in models like GPT~\cite{radford2018improving}, is optimized for autoregressive generation, making it suitable for text generation, code synthesis, and other generative tasks. The encoder–decoder design, seen in models such as T5~\cite{raffel2020exploring}, combines both components to map input sequences to output sequences, supporting tasks like translation, summarization, and multimodal generation. These flexible designs enable Transformer to serve as a general backbone for a wide range of FMs. In addition to Transformer, the diffusion model~\cite{ho2020denoising} has become a powerful architectural paradigm for \glspl{fm}, particularly in generative tasks involving images, audio, and multimodal content. Diffusion models operate through a two-stage process: progressively corrupting data with noise (forward diffusion) and training a neural network to reverse this process through iterative denoising. A key strength of diffusion models lies in their ability to produce high-fidelity and diverse samples with stable training, in contrast to earlier generative paradigms such as generative adversarial networks (GANs)~\cite{goodfellow2020generative}. This capability has enabled state-of-the-art text-to-image \glspl{fm} such as DALLE~\cite{ramesh2021zero}, Stable Diffusion~\cite{esser2024scaling}, and Midjourney~\cite{midjourney2025}.

\glspl{fm} can be categorized based on their modality and functionality. 
\glspl{llm}, a typical type of \gls{fm}, are trained primarily on large text datasets and excel in natural language understanding and generation. They support a wide range of tasks, including text classification, summarization, machine translation, conversational AI, and code generation. 
Vision \glspl{fm} are trained on massive and diverse image datasets to capture rich semantic and structural visual representations that can be adapted to various vision tasks. Vision \glspl{fm} can be built on different architectures, such as encoder-only models like Vision Transformers~\cite{dosovitskiy2020image} for image classification, encoder–decoder models for tasks such as visual question answering, or diffusion-based models for high-fidelity image synthesis. Multimodal \glspl{fm} are trained to process multiple types of data, such as text, images, audio, and video, with a single \gls{fm}. They often combine modality-specific encoders (e.g., a text transformer and a vision transformer) with shared embedding spaces or cross-attention mechanisms for fusion to learn aligned representations across different modalities. Multimodal \glspl{fm} are capable of reasoning about relationships between different kinds of inputs and therefore can perform a variety of cross-domain tasks, such as aligning images with textual descriptions, generating images from text prompts, or producing video summaries from audio–visual input.

\subsection{Cyber-physical Systems and their Software}

\begin{figure}[ht]
    \centering
    \includegraphics[width=0.7\linewidth]{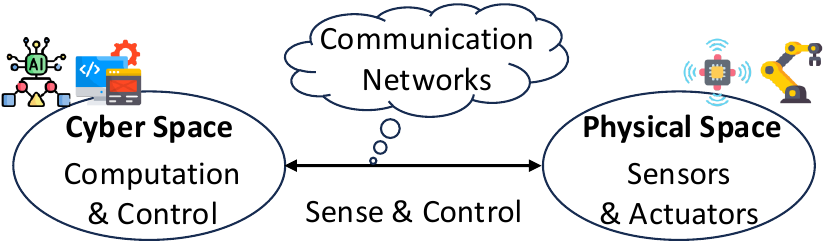}
    \caption{Architecture of Cyber-physical Systems}
    \label{fig:cps_arch}
    \vspace{-6pt}
\end{figure}

Cyber-physical systems (CPSs)~\cite{baheti2011cyber,CPSDomainAnalysis} refer to systems that integrate computational and physical capabilities, continuously interacting with the physical world through computation, communication, and control. As Figure~\ref{fig:cps_arch} shows, a typical CPS consists of a computation and control unit, a set of sensors for data acquisition, actuators for interacting with the physical environment, and a communication infrastructure for transmitting data between the physical and cyber domains. These components coordinate to enable CPS functions to operate autonomously and adaptively in a dynamic environment, allowing CPSs to monitor, respond to, and adapt to changes in real-time. Specifically, the control unit serves as the core of a CPS, which processes sensor data, executes control or decision-making algorithms, and issues commands to actuators. Sensors capture the current state of the physical world (e.g., speed, temperature, and camera view), which is then processed by the control unit to make informed decisions. Actuators execute these decisions and physically interact with the environment by taking actions such as moving a robotic arm, adjusting the valve, or controlling vehicle throttles. The communication infrastructure ensures reliable and timely data exchange between sensors, actuators, and the control unit, and, in some cases, with distributed computing nodes or cloud services.

CPSs have become essential technologies for transforming a wide range of industrial applications, particularly in the context of Industry 4.0~\cite{jazdi2014cyber}, which emphasizes smart factories, interconnected production systems, and data-driven automation. Key domains include autonomous vehicles, industrial automation, robotics, aerospace, healthcare, and smart infrastructure. In these applications, CPSs must satisfy strict requirements for real-time performance, reliability, safety, and security, while handling heterogeneous and dynamic environments. The integration of computation, communication, and control in CPSs enables adaptive, intelligent, and efficient industrial operations, but also introduces significant challenges in system design, software development, and validation.

Software plays a pivotal role in CPS functionality, serving as the decision-making agent that connects sensing, computation, and actuation. Different from traditional software systems, CPSs integrate software with physical processes, requiring continuous interaction with the real world through sensors and actuators. They must handle heterogeneous, multimodal data and operate in dynamic and uncertain environments, while often being constrained by strict real-time and synchronization requirements. Moreover, CPSs demand close integration across software, hardware, and networking, making robustness and assurance central concerns. Beyond traditional software system design challenges, such as maintainability and scalability, CPS software presents unique challenges related to real-time performance, safety-critical operation, heterogeneity of hardware and communication protocols, and integration with dynamic physical environments. For instance, the heterogeneity of hardware and communication protocols introduces additional complexity, as software must seamlessly coordinate components with differing capabilities, interfaces, and timing characteristics. Furthermore, integration with dynamic physical environments necessitates adaptive and robust software architectures that can respond to changing conditions, uncertainties, and external disturbances while maintaining system stability and desired functionality. These challenges distinguish CPS software from conventional software systems and necessitate specialized design, testing, and assurance methodologies.
In recent years, \glspl{fm}, particularly multi-modal ones, have attracted growing attention in the field of \gls{cps}, as these often operate in complex, dynamic, and heterogeneous environments where information comes from multiple modalities. However, current practices primarily focus directly on integrating \glspl{fm} into \gls{cps} operation (e.g., NVIDIA's GR00T-N1~\cite{bjorck2025gr00t}). Regarding software engineering for \gls{cps}, \glspl{fm} have the potential to transform several key stages of the development lifecycle. For instance, in verification and validation, multimodal \glspl{fm} can process system logs, sensor data, and execution traces to detect anomalies, predict failures, and enhance testing coverage. Finally, at runtime, adaptive \glspl{fm} can enable \gls{cps} to respond to unforeseen environmental changes by interpreting multimodal inputs and guiding real-time decision-making. In this paper, we foresee significant opportunities for applying \glspl{fm} in software engineering for \gls{cps}, ranging from automated requirements analysis and code generation to verification, validation, and adaptive runtime support. At the same time, we also identify key challenges when applying \glspl{fm} into \gls{cps} software, including handling data heterogeneity, ensuring safety and reliability assurance, improving model interpretability, and integrating \glspl{fm} into existing \gls{cps} software lifecycle.

\revision{
\section{Roadmap Methodology and Overview}\label{sec:method}

\subsection{Methodology}
As shown in Figure~\ref{fig:methodology}, we follow an iterative process to develop this research roadmap.
The process begins with \circlednum{1} \textit{Initial Discussion \& Brainstorming}, where we follow an approach inspired by the Delphi method~\cite{linstone1975delphi} and the nominal group technique~\cite{harvey2012nominal}. The Delphi method gathers and refines expert opinions through iterative rounds of individual input and collective feedback to converge toward consensus, while the nominal group technique complements this by structuring group discussions such that participants first generate ideas independently before presenting and deliberating them collectively, preventing dominant voices from suppressing minority perspectives. In our context, each author first independently identifies research opportunities and key challenges based on their expertise and familiarity with the relevant literature. Authors then collectively discuss and extend these points through structured group sessions to reach consensus on the final set of opportunities and challenges to be included in the roadmap. Each author subsequently writes and refines their assigned draft.
The draft then undergoes \circlednum{2} \textit{Internal Validation I}, in which senior authors cross-validate each section for clarity, consistency, and correctness. The validated roadmap is then submitted for \circlednum{3} \textit{Discussion at SE2030 Workshop} (co-located with FSE 2025, Trondheim, Norway), where ideas are shared and discussed with external participants from diverse domains using liberating structures, including flash keynotes, impromptu networking, fishbowl, and other activities. Based on the workshop outcomes, \circlednum{4} \textit{Integration \& Extension} is performed, where authors incorporate participant feedback, revise arguments, and extend the roadmap content. The extended paper then undergoes \circlednum{5} \textit{Internal Validation II}, a final comprehensive review, and quality assurance pass by the senior authors. Finally, the \circlednum{6} \textit{FM4CPS Roadmap} is submitted to the ACM TOSEM 2030 Roadmap for Software Engineering.

\begin{figure}[ht]
    \centering
    \includegraphics[width=.98\linewidth]{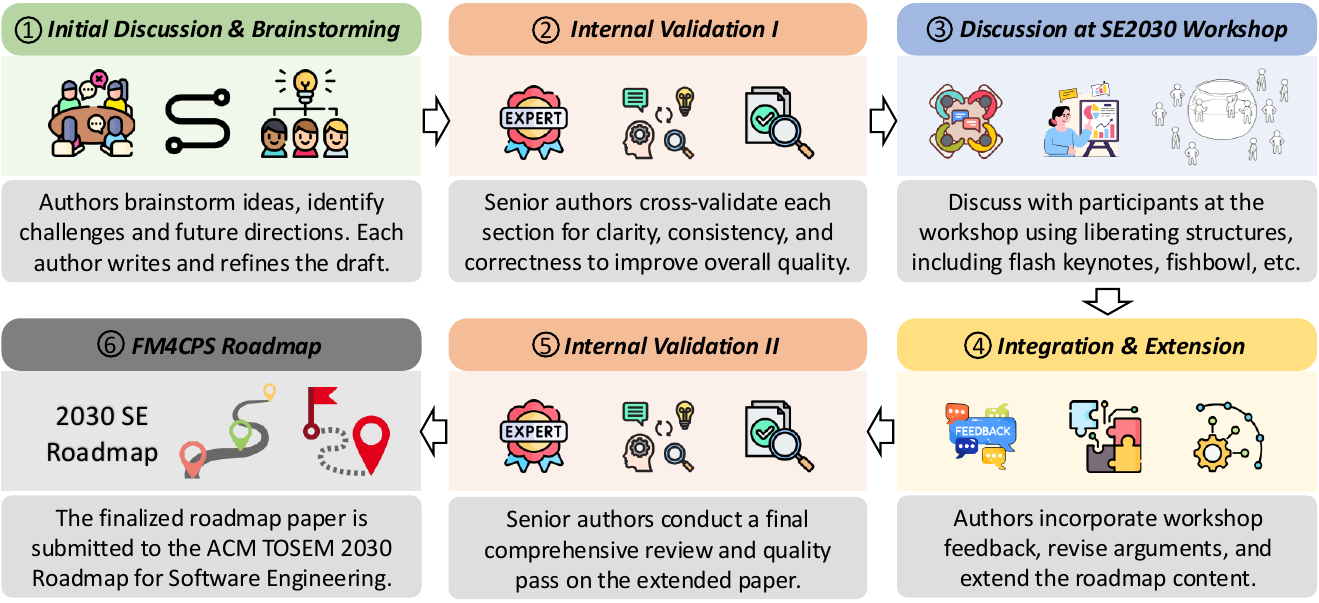}
    \caption{Overview of the Roadmap Development Process}
    \label{fig:methodology}
\end{figure}

Notice that this paper is intended as a visionary, forward-looking guide rather than a systematic literature review; therefore, we do not follow the procedures of systematic literature reviews. However, where possible, we support our arguments with references to the existing literature to strengthen the roadmap's foundation. As a research roadmap, its primary aim is to present representative challenges and opportunities, guide researchers, and direct future research.

\subsection{Roadmap Overview}
Figure~\ref{fig:challenges_opportunities} presents an overview of the challenges and research opportunities identified in this roadmap, covering the entire \gls{cps} software lifecycle across six development phases, as well as a set of common cross-cutting challenges. The six phases cover \textit{Requirements Engineering} (Section~\ref{sec:RE4CPS}), \textit{Design and Modeling} (Section~\ref{sec:Model4CPS}), \textit{Software Development} (Section~\ref{sec:SoftwareDev4CPS}), \textit{Testing} (Section~\ref{sec:CPSTest}), \textit{Debugging and Repair} (Section~\ref{sec:CPSrepasir}), and \textit{Evolution} (Section~\ref{sec:CPSevolution}), while the common cross-cutting challenges spanning technical, safety and certification, economic and resource, human and organizational, ethical and privacy, and environmental dimensions are discussed in Section~\ref{sec:challenges}.



\begin{figure}[ht]
    \centering
    \includegraphics[width=0.85\textwidth]{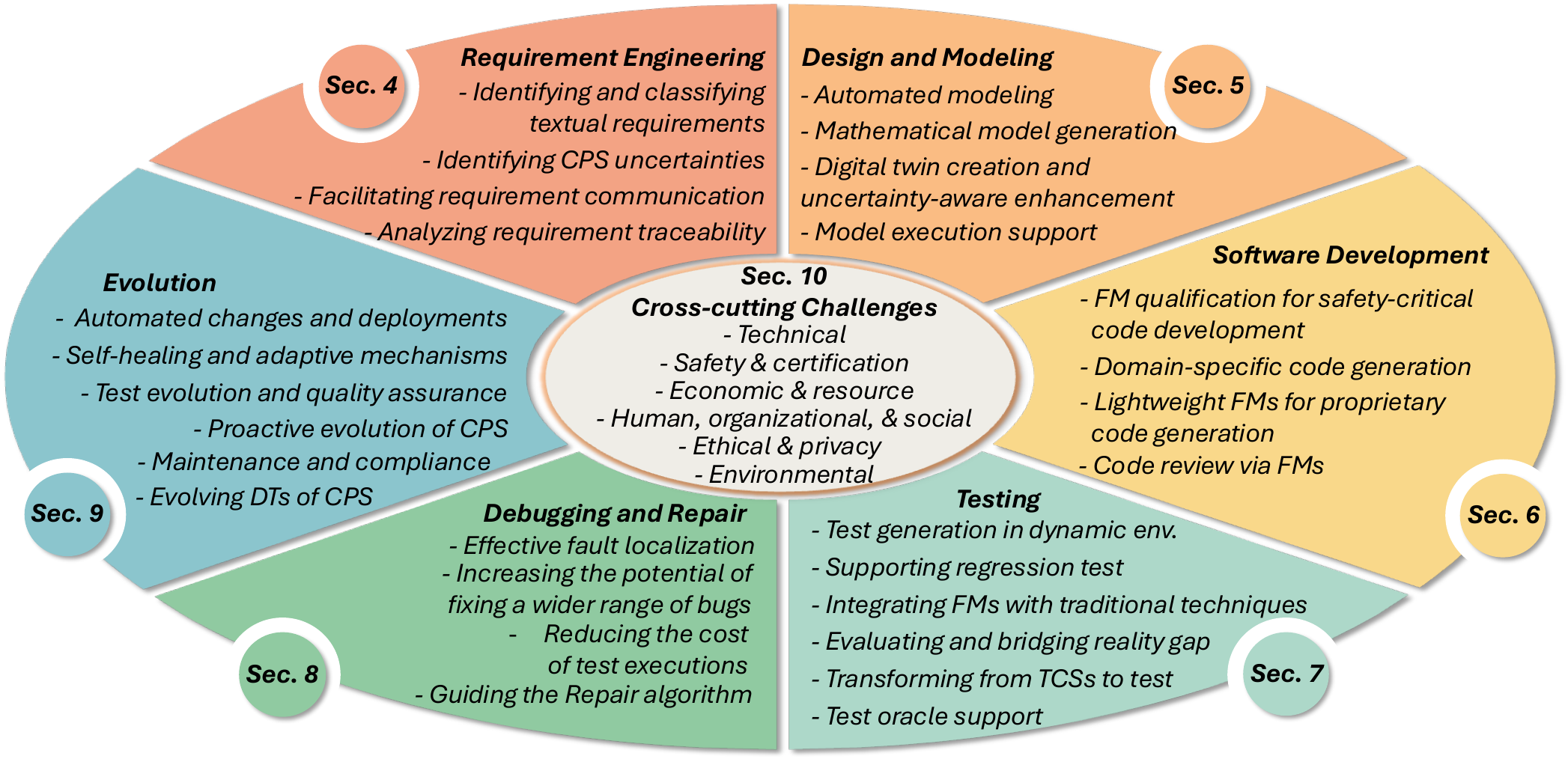}
    \caption{Challenges and Opportunities of \glspl{fm} for Software Engineering of \glspl{cps}}
    \label{fig:challenges_opportunities}
\end{figure}



%
%
For each \gls{cps} software development phase, we identify challenges and outline corresponding research opportunities formulated as key research questions. We further suggest methodologies to address these questions, along with baseline approaches and evaluation metrics, to enable empirical evaluation of the cost-effectiveness of \glspl{fm}-based solutions.
}

\section{Foundation Models for Requirements Engineering} \label{sec:RE4CPS}
\revision{This section discusses how \glspl{fm} can potentially support requirements engineering workflows in \gls{cps} development. We first provide, in Section \ref{subsec:re-backsota}, a brief background and overview of the state of the art regarding the use of \glspl{fm} for \gls{cps} requirements engineering. Following that, Section \ref{subsec:re-challengedandRo} presents the identified challenges and shows how these challenges result in relevant research questions, along with possible methodologies to address them.}

















%

\subsection{Background and State of the Art} \label{subsec:re-backsota}
\Gls{re} is the process of identifying, analyzing, documenting, and validating the descriptions of a system's functionality, services, and operational constraints in accordance with stakeholder needs~\cite{sommerville2011software}. 
For \glspl{cps}, which integrate computational processes with physical components operating in dynamic environments, these systems must satisfy stringent operational and reliability requirements to function safely. 
The \gls{re} process for \glspl{cps} is particularly challenging because their architectures integrate hardware, software, sensors, actuators, and communication networks, requiring requirements that address both cyber and physical aspects and supporting modular specification and validation~\cite{wiesner2014requirements}. 
Given the safety-critical nature of many \gls{cps} applications, their requirements must capture real-time constraints and account for uncertainties such as sensor failures, network errors, and anomalous behavior~\cite{khayatian2022plan}. 
In addition, the involvement of diverse stakeholders and the interdisciplinary development of \glspl{cps} further complicate the process of eliciting, prioritizing, and validating requirements~\cite{robinson2021bridging}.

\gls{re} for \gls{cps} research covers aspects, e.g., requirements elicitation, formal specification, modeling, and verification~\cite{ayerdi2020towards,nuzzo2018chase,bouskela2022formal}. 
Several studies exist for security requirements analysis to ensure \gls{cps} resilience against threats~\cite{nguyen2017model,wach2020model}. 
Moreover, some works focus on requirements-driven \gls{cps} development~\cite{Vinarek_2016} and on uncertainty modeling to enhance \glspl{cps} adaptability and robustness~\cite{ahmad2018towards}. 
\glspl{llm} have been recently used to improve several \gls{re} tasks, e.g., checking requirements completeness, requirements classification, and requirement translation~\cite{peer2024nlp4ref,norheim2024challenges}. 
However, using \glspl{fm} for \gls{re} for \gls{cps} remains unexplored, opening up new research opportunities. \looseness=-1 

\subsection{Challenges and Research Opportunities} \label{subsec:re-challengedandRo}
\revision{In this section, we provide a list of challenges related to the application of applying \glspl{fm} to requirements engineering of \gls{cps} (see Table \ref{tab:cps-re-challenges}). These challenges are identified by reading the related literature discussed in Section \ref{subsec:re-backsota}. For each challenge, we provide research opportunities, together with research questions, relevant methodologies to apply, and baselines and evaluation metrics to assess the cost-effectiveness of these methodologies. The overall mapping between the challenges and research opportunities is summarized in Table \ref{tab:re-research-opportunities}.   
}

\begin{table*}[t]
{\color{black}
\centering
\small
\caption{\revision{Challenges in Applying Foundation Models to CPS Requirements Engineering}}
\label{tab:cps-re-challenges}
\rowcolors{2}{gray!20}{white}
\begin{tabular}{p{0.15\textwidth} p{0.8\textwidth}}
\rowcolor{black!70}
\textcolor{white}{\textbf{Challenge ID}} &
\textcolor{white}{\textbf{Challenge Description}} \\

$\mathbf{Ch}_{\mathbf{RE1}}$ &
Ambiguity, incompleteness, and contradictions in natural-language CPS requirements hinder automated interpretation and downstream analysis. \\

$\mathbf{Ch}_{\mathbf{RE2}}$ &
Processing and verifying multimodal requirements artifacts (e.g., text, diagrams, tables) remains error-prone and weakly supported by current tooling. \\

$\mathbf{Ch}_{\mathbf{RE3}}$ &
Identifying, modeling, and managing uncertainty sources and their operational impact in CPSs remains difficult and insufficiently supported. \\

$\mathbf{Ch}_{\mathbf{RE4}}$ &
Establishing and maintaining traceability between requirements and CPS development artifacts requires substantial manual effort and does not scale. \\

$\mathbf{Ch}_{\mathbf{RE5}}$ &
Classifying and evolving functional and non-functional requirements over time is costly and complicates prioritization and impact analysis. \\

$\mathbf{Ch}_{\mathbf{RE6}}$ &
Missing or imprecise requirements and communication gaps among heterogeneous stakeholders reduce requirement quality and consistency. \\

$\mathbf{Ch}_{\mathbf{RE7}}$ &
Transforming textual requirements into downstream software engineering artifacts remains largely manual and error-prone. \\

\end{tabular}
}
\end{table*}





\begin{table}[htbp]
{\color{black}
\centering
\footnotesize
\caption{\revision{Actionable Research Opportunities Corresponding to the Identified Challenges for \gls{cps} Requirements Engineering with Foundation Models}}
\label{tab:re-research-opportunities}
\begin{tabular}{|p{0.05\textwidth}|p{0.17\textwidth}|p{0.30\textwidth}|p{0.15\textwidth}|p{0.20\textwidth}|}
\hline
\rowcolor{black!70}
\color{white}\textbf{ID.} & \color{white}\textbf{Research Question} & \color{white}\textbf{Methodology} & \color{white}\textbf{Baseline(s)} & \color{white}\textbf{Evaluation Metrics} \\
\hline
\rowcolor{white}
$\mathrm{Ch}_{\mathrm{RE1}}$ & How can \glspl{llm} identify, resolve, and improve contradictory or uncertain textual requirements? & Fine-tune \glspl{llm} on \gls{cps} requirement specification datasets belonging to a domain (e.g., autonomous driving system) and then use them to detect inconsistencies, ambiguities, and gaps; suggest improvements to experts. & Natural language processing, Model-based techniques & Precision/recall/F1 scores in identifying and detecting contradictory requirements, reduction in manual effort for this same task, improvement in supporting requirements engineers in clarifying the requirements. \\
\rowcolor{gray!20}
$\mathrm{Ch}_{\mathrm{RE2}}$& How can multimodal \glspl{fm} verify, classify, and process requirements expressed in mixed-mode formats? & Use prompt engineering to fine-tune multimodal \glspl{fm} with textual, diagrammatic, and sensor requirements (including audio and video); perform semantic and structural consistency checks. & Deep learning methods, model validation techniques, Formal methods & Accuracy, Time, Robustness, Resource utilization \\
\rowcolor{white}
$\mathrm{Ch}_{\mathrm{RE3}}$ & How can \glspl{llm}-based tools systematically identify and manage uncertainties at different \gls{cps} development stages? & Develop advanced prompting techniques utilizing \gls{cps} uncertainty taxonomies, \gls{cps} artifacts (e.g., requirements and code), and expert guidance to iteratively identify and manage uncertainties. & Formal modeling, Monitoring and analysis methods, Testing techniques, Manual identification by experts & Accuracy of uncertainty predictions, coverage of uncertain scenarios, detection of unknown scenarios, and risk quantification. \\
\rowcolor{gray!20}
$\mathrm{Ch}_{\mathrm{RE4}}$ & How can \glspl{fm} facilitate automated traceability throughout the \gls{cps} development lifecycle? & Utilize few-shot active learning with \glspl{fm} by using requirements-artifact pairs (e.g., code, models, tests) from a specific \gls{cps} domain and suggest traceability links to experts for validation. & Deep learning, Natural language processing, Model-driven methods & Precision/recall/F1 scores, reduction in manual effort, alignment with traceability established by experts. \\
\rowcolor{white}
$\mathrm{Ch}_{\mathrm{RE5}}$ & How can \glspl{fm} classify, prioritize, and align \gls{cps} requirements based on criticality, environmental factors, and stakeholder needs? & Fine-tune \glspl{fm} on labeled requirements datasets with metadata on criticality, environmental factor, and context where available; implement automated classification, prioritization, and alignment suggestions. & Reinforcement learning, Search-based methods, Domain-specific modeling & Classification accuracy, prioritization effectiveness, stakeholder satisfaction. \\
\rowcolor{gray!20}
$\mathrm{Ch}_{\mathrm{RE6}}$ & How can \glspl{llm} identify missing requirements and enhance the quality of existing ones? & Apply \glspl{llm} with prompting engineering techniques to analyze requirement documents and project communications to detect gaps or incomplete specifications; suggest refinements. & Natural language processing, Validation techniques, Knowledge graphs & Number of defects detected, reduction in requirement omissions, improvement in requirement quality compared to the one performed by an expert. \\
\rowcolor{white}
$\mathrm{Ch}_{\mathrm{RE7}}$ & How can \glspl{fm} automate the generation of SE artifacts from multi-formatted \gls{cps} requirements? & Train \glspl{fm} on paired datasets of requirements and corresponding artifacts (models, test cases, code templates); generate candidate artifacts with validation support. & Natural language processing, Search algorithms, Model transformation methods & Accuracy and completeness of generated artifacts, reduction in manual effort. \\
\hline
\end{tabular}
}
\end{table}

\subsubsection{Handling Requirements Elicitation Challenges with \glspl{fm}}
High-level \gls{cps} requirements are often elicited in natural language for stakeholder communication. 
Natural language requirements can be ambiguous, incomplete, or contradictory. 
Therefore, many works support requirements specifications using formal and modeling languages, which require domain experts to have expertise in such languages. 
In addition, requirements can be specified in different formats that involve multimodal data such as user interfaces, images, and videos. 
\revision{However, traditional requirement engineering techniques were not designed to process diverse and multimodal data formats~\cite{pei2022requirements}.} 
Given these challenges, we see \glspl{fm}' potential in many aspects. 
First, \glspl{llm} with textual analysis and comprehension capabilities can \revision{use \gls{cps} requirements documents to identify uncertain and contradictory requirements~\cite{fantechi2023inconsistency}.} 
Second, for \gls{cps} requirements specified in multiple formats, \revision{multimodal \glspl{fm} with mixed-mode capabilities can be used to verify requirement specifications such as formal models, user interfaces, and design models~\cite{hamalainen2025systematic}.}
Third, \revision{more research is needed to investigate using \glspl{llm} to identify requirements during \gls{cps} development and improve the quality of existing \gls{cps} requirements that may be imprecise~\cite{zadenoori2025llms}.} 
Finally, to facilitate easier communication and prevent miscommunication among diverse stakeholders, \revision{\glspl{fm} hold great potential to tailor the same set of requirements to match the backgrounds of different stakeholders~\cite{zadenoori2025llms}.} 
For example, they can generate user interface designs from textual requirements or vice versa. 
However, more research is needed to explore how \glspl{fm} can improve communication among stakeholders involved in \gls{cps} development.

\subsubsection{Identifying Uncertainties in \gls{cps} from Requirements}
\Glspl{cps} face uncertainties in their operation, e.g., due to environmental conditions, sensor inaccuracies, and human interactions. 
A key challenge is to identify the sources and impacts of uncertainties in \glspl{cps}. 
To identify uncertainties, domain experts typically rely on their instincts and past experiences~\cite{sartaj2024uncertainty,sartaj2025identifying}. 
Given LLMs' text comprehension and analytical capabilities, \revision{domain experts can use \gls{cps} requirements along with advanced prompting techniques to systematically identify uncertainties at various stages of development~\cite{arora2024advancing}.} 
A recent study reported LLMs' potential in identifying uncertainties in self-adaptive robotics using their requirements~\cite{sartaj2025identifying}. 
Beyond that, we foresee the need to develop LLM-based approaches and tools to assist domain experts in uncertainty identification. 
Moreover, some \gls{cps} requirements can also be specified in different formats, \revision{such as Simulink models with visual aspects~\cite{luitel2024requirements}.} 
Therefore, \revision{a potential future direction is to develop new techniques based on multimodal \glspl{fm} to identify uncertainty using multi-format data~\cite{song2024multi}.} 

\subsubsection{Requirements Traceability Analysis with \glspl{fm}}
Requirements traceability links \gls{cps} requirements with various artifacts in the development phases, such as software models, source code, and tests. 
\revision{Establishing traceability in \glspl{cps} often requires manual traceability analysis~\cite{mucha2024systematic}.} 
To this end, future research can focus on two directions. 
First, using \glspl{llm}' natural language capabilities to automatically identify, establish, and update traceability links from \gls{cps} requirements to the development artifacts, \revision{such as code and test cases in textual format~\cite{dearstyne2025intelligent}.}
Second, using multimodal \glspl{fm} for requirements to ensure traceability across the entire \gls{cps} development lifecycle, \revision{including models, code, and \gls{cps} simulations~\cite{song2024multi}.}

\subsubsection{Requirements Classification with \glspl{fm}}
\Gls{cps} requirements typically include diverse aspects, including functional and non-functional requirements and constraints among them. 
\revision{Classifying requirements for \glspl{cps} is challenging due to their continuous evolution and the complex nature of these systems~\cite{schneider2018requirements}.} 
In this regard, a potential research direction involves using \glspl{llm} to assist domain experts in selecting and prioritizing requirements based on their criticality level while also considering aspects, \revision{such as environmental factors~\cite{hassan2024rethinking}.}
Another research direction is to explore VLMs and multimodal \glspl{fm} for \revision{classifying requirements modeled with graphical languages~\cite{yang2024foundation}.}

\subsubsection{Identifying Missing Requirements and Improving Existing Ones with \glspl{fm}} When developing a \gls{cps}, the complete set of textual requirements under which it shall operate is either unknown or imprecise, \revision{which could eventually lead to incorrect implementation~\cite{kasauli2021requirements}.} To this end, more research is needed to investigate the use of \glspl{llm} for identifying requirements during \gls{cps} development, \revision{as well as for improving the quality of existing \gls{cps} requirements that may be imprecise~\cite{cheng2025generative}.} This is based on the assumption that \glspl{llm} are trained on vast amounts of knowledge and could assist requirements engineers with requirements engineering activities.

Moreover, many \glspl{cps} work with different data modalities. For instance, \glspl{av} use various sensors to identify their environment. Thus, \revision{there is potential to automatically extract textual requirements for a \gls{cps} from historical data available in different formats to support the requirements engineering of \glspl{cps}~\cite{aluvalu2025cyber}.} In our example, based on past videos from traffic cameras, one can extract requirements on the operating environment of \glspl{av} in textual format.

\subsubsection{Facilitating Requirements Communication Among Diverse Stakeholders} 
\glspl{cps}, being interdisciplinary in nature, involve diverse stakeholders. \revision{Communication among them has always been a significant challenge that impacts the quality of the \gls{cps} being developed~\cite{cederbladh2024experiences}.} To this end, \revision{\glspl{fm} hold great potential to tailor the same set of requirements to match the backgrounds of different stakeholders, facilitating easier communication and preventing miscommunication~\cite{gallaba2025conversational}.} However, further research is needed to explore how \glspl{fm} can enhance communication among stakeholders involved in \gls{cps} development.   

\subsubsection{Textual Requirements to Other Software Engineering Artifacts} 
Commonly, \revision{the requirements from the requirements engineering phase are manually converted into other software engineering artifacts, such as design models, code, and tests~\cite{sundaram2010assessing}.} With \glspl{fm}, \revision{there is potential to translate textual requirements directly into these artifacts~\cite{rodriguez2023prompts}.} However, this research area for \glspl{cps} remains understudied, and there are ample research opportunities to build customized \glspl{fm} to support the translation of requirements into models, code, and tests.

\section{Foundation Models for System Design and Modeling} \label{sec:Model4CPS}
\revision{This section explores the role of \glspl{fm} in supporting system design and modeling throughout the development of \glspl{cps}. An overview of the necessary background and current state of the art is provided in Section~\ref{subsec:SD_back}. Subsequently, Section~\ref{subsec:SD_cro} analyzes the main challenges encountered in practice and derives corresponding research questions, outlining potential directions for addressing these challenges.}












\revision{
\subsection{Background and State of the Art}\label{subsec:SD_back}

Software design and modeling play a crucial role in \gls{cps} development, forming the foundation for system standards and lifecycle management~\cite{derler2011modeling}. In the era of \glspl{fm}, researchers have leveraged \glspl{fm} to support \gls{cps} software design and modeling~\cite{timperley2025assessment,wang2024llms,ferrari2024model,camara2023assessment}, 
while shortcomings remain in critical aspects, such as syntactic and semantic correctness, consistency, and scalability.
Moreover, the generation of mathematical models based on \glspl{fm} is a promising possibility, given the increasing mathematical reasoning and data analysis capabilities of \glspl{fm}~\cite{imani2023mathprompter,azerbayev2023llemma,luo2023wizardmath,jansen2025leveraging}. However, as Ahn et al.~\cite{ahn2024large} pointed out, existing \glspl{fm} still struggle with limited generalization in mathematical problems and exhibit weaknesses in mathematical reasoning.
Furthermore, \gls{dt}, as an innovative design approach, has attracted widespread attention from researchers, and the powerful knowledge representation and reasoning capabilities of \glspl{fm} are actively driving the rapid development of \glspl{dt}~\cite{trantas2024digital,huang2025generative,ali2024foundation}. However, current \gls{fm} technology has limitations and cannot fully support an \gls{fm}-based \gls{dt} solution~\cite{ali2024foundation}.
Additionally, \glspl{fm} can support model execution by generating action code, evaluating runtime constraints, and synthesizing input data for simulation and testing~\cite{ren2025simugen,xiang2025modigen,lu2025assessing}. However, existing approaches still have limited support for complex dependencies and system-level coherence, indicating that further exploration in this direction remains necessary~\cite{xiang2025modigen}.
}

\subsection{Challenges and Research Opportunities}\label{subsec:SD_cro}
\revision{This section presents an overview of key challenges in the application of \glspl{fm} to \gls{cps} software design and modeling, derived from the literature surveyed in the preceding section, as shown in Table~\ref{tab:design-modeling-challenges}. For each challenge, we outline associated research opportunities, research questions, and potential methodologies, as well as baselines and metrics for assessing the cost-effectiveness of these methodologies. The relationships between challenges and research opportunities are summarized from different aspects in Table~\ref{tab:modeling-opportunities},~\ref{tab:fm-math-model-opportunities},~\ref{tab:fm-dt-opportunities}, and~\ref{tab:fm-model-execution-opportunities}.}

\begin{table*}[t]
{\color{black}
\centering
\small
\caption{\revision{Challenges in Applying Foundation Models to \gls{cps} System Design and Modeling}}
\label{tab:design-modeling-challenges}
\rowcolors{2}{gray!20}{white}
\begin{tabular}{p{0.15\textwidth} p{0.8\textwidth}}
\rowcolor{black!70}
\textcolor{white}{\textbf{Challenge ID}} &
\textcolor{white}{\textbf{Challenge Description}} \\

$\mathbf{Ch}_{\mathbf{DM1}}$ &
Handling heterogeneous and ambiguous input data, which can lead to inconsistent models. \\

$\mathbf{Ch}_{\mathbf{DM2}}$ &
Constructing physics-grounded, domain-specific models capturing complex CPSs as well as their environmental behavior. \\

$\mathbf{Ch}_{\mathbf{DM3}}$ &
Reducing expert dependency and mitigating error-prone workflows in CPS and environment modeling. \\

$\mathbf{Ch}_{\mathbf{DM4}}$ &
Addressing dynamic system evolution and real-time constraints, which require adaptive modeling solutions. \\

$\mathbf{Ch}_{\mathbf{DM5}}$ &
Ensuring model fidelity, generalization, and validation to guarantee model correctness and faithfully reflect CPSs and their environment. \\

\end{tabular}
}
\end{table*}


\subsubsection{\gls{fm}-based \gls{cps} Modeling}\label{subsubsec:fm4cpsmodeing}

\begin{table}[htbp]
{\color{black}
\centering
\small
\caption{\revision{Actionable Research Opportunities Corresponding to the Identified Challenges for \gls{cps} Modeling with Foundation Models (Section~\ref{subsubsec:fm4cpsmodeing})}}
\label{tab:modeling-opportunities}
\renewcommand{\arraystretch}{1.3} 
\begin{tabular}{|p{0.06\textwidth}|p{0.20\textwidth}|p{0.27\textwidth}|p{0.15\textwidth}|p{0.20\textwidth}|}
\hline

\rowcolor{black!70}
\color{white}\textbf{ID.} & \color{white}\textbf{Research Question} & \color{white}\textbf{Methodology} & \color{white}\textbf{Baseline(s)} & \color{white}\textbf{Evaluation Metrics} \\
\hline
\rowcolor{white}
$\mathrm{Ch}_{\mathrm{DM1}}$ & How can domain-specific fine-tuning improve \glspl{fm} support for multi-modal, context-rich \gls{cps} modeling dialogues? & Develop and fine-tune \glspl{fm} on diverse annotated \gls{cps} modeling dialogues, such as across text, diagrams, and structured data. & \gls{fm} without fine-tuning, Deep learning, Non-ML. & Accuracy of recommendations, completeness of modeling steps, and user satisfaction. \\
\rowcolor{gray!20}
$\mathrm{Ch}_{\mathrm{DM3}}$ & How can memory architectures in \glspl{fm} retain long-term modeling context and track evolving system representations? & Design memory mechanisms within \glspl{fm} to incorporate long-term context retention, cross-modal fusion across modeling sessions. & \glspl{fm} without memory augmentation; & Modeling consistency across different sessions, accuracy in recalling previously defined components, reduction in user corrections and repetitive modeling tasks. \\
\rowcolor{white}
$\mathrm{Ch}_{\mathrm{DM1}}$ & How can multimodal inputs be interpreted consistently by \gls{fm}-based modeling assistants? & Develop semantic alignment techniques to process text, diagrams, and structured inputs into consistent semantic representations. & Manual, single-modal \glspl{fm}; & Semantic consistency across modalities, alignment with manually created models, and reduction in interpretation errors. \\
\rowcolor{gray!20}
$\mathrm{Ch}_{\mathrm{DM3}}$, $\mathrm{Ch}_{\mathrm{DM4}}$ & How can \gls{cps} modeling environments provide real-time, context-aware \gls{fm} guidance while enforcing constraints? & Integrate \glspl{fm} into interactive modeling environments, implement constraint-checking modules, and provide context-aware suggestions. & Non-ML, Static rule-based modeling assistants; & Response time, guidance relevance, constraint satisfaction, rate of constraint 
violations detected, efficiency. \\
\rowcolor{white}
$\mathrm{Ch}_{\mathrm{DM3}}$, $\mathrm{Ch}_{\mathrm{DM5}}$ & How can \glspl{fm} automate partial model completion and code generation in diverse \gls{cps} modeling environments? & Fine-tune \glspl{fm} on paired datasets of partial models and completed artifacts/code, develop auto-completion functionality. & Deep learning, Natural language processing; & Accuracy of generated model components, correctness of completed model components and code, reduction in manual effort. \\
\rowcolor{gray!20}
$\mathrm{Ch}_{\mathrm{DM1}}$, $\mathrm{Ch}_{\mathrm{DM2}}$ & How can end-to-end \gls{fm} pipelines translate natural language \gls{fm} requirements into structural and behavioral models? & Build pipelines that convert textual requirements into model templates, validated through simulation or formal analysis. & Manual model construction, model transformation tools; & Correctness and completeness of generated models, coverage of requirements. \\
\rowcolor{white}
$\mathrm{Ch}_{\mathrm{DM2}}$, $\mathrm{Ch}_{\mathrm{DM5}}$ & How can human-in-the-loop mechanisms ensure realism, hybrid system correctness, and functional alignment in generated models? & Develop interactive FM interfaces where experts can validate, refine, or constrain generated models. & automated model generation without human feedback, manual model construction; & Reduction in expert interventions, model correctness, alignment with intended functionality, and user satisfaction. \\
\hline
\end{tabular}
}
\end{table}

Modeling CPS is highly challenging due to its multidisciplinary nature, where different domains have distinct modeling abstractions, semantics, and design constraints~\cite{derler2011modeling}. The modeling process is further complicated by heterogeneous and multimodal inputs (e.g., natural language requirements, system diagrams, control logic, and time-series sensor data), which are difficult to integrate using traditional modeling tools~\cite{baris2025foundation}. The modeling process is often manual, expert-dependent, and error-prone, leading to high costs, low consistency, and limited reusability~\cite{alenazi2019sysml}.
\glspl{fm}, with their capabilities in language understanding, cross-modal reasoning, and knowledge integration, have the potential to alleviate these challenges. The following sections explore their roles and advantages in interactive modeling, embedded workflow support, and end-to-end model generation.

\textbf{Interactive Modeling via FM-based Chatbot.} 
\glspl{fm} can act as interactive modeling assistants through chatbot-based interfaces, supporting users via flexible and multimodal interactions combining textual queries, system diagrams, parameter tables, partial models, and simulation traces.
This allows modelers to express design intent and uncertainty in richer and more natural ways across different modeling contexts.
In exploratory cases, \gls{fm}-based chatbot can interpret ambiguous requirements, identify candidate components, and suggest initial modeling patterns. 
In task-specific modeling phases, where the overall system goals are defined but implementation details remain open, the chatbot assistant can assist with more focused design questions. For instance, \glspl{fm} can analyze the partial input and provide design recommendations grounded in relevant system constraints. Through such interactions, \glspl{fm} enable iterative refinement and continuous support throughout the modeling lifecycle.



To improve the effectiveness of multimodal interactions, several research challenges must be addressed. In particular, domain-specific fine-tuning on annotated modeling dialogues that integrate text, diagrams, and structured data is essential, as prompt engineering alone is insufficient for handling context-rich, multimodal dialogue scenarios. Parameter-efficient methods, such as adapter tuning~\cite{houlsby2019parameter}, Low-Rank Adaptation (LoRA)~\cite{hu2022lora}, and prefix tuning~\cite{li2021prefix}, provide practical approaches to align \glspl{fm} with domain-specific reasoning and terminology.
Moreover, maintaining context across dialogue turns is essential. FMs must be able to remember previously defined components, unresolved design choices, and user intentions throughout the extended modeling process. However, while recent memory enhancement techniques, including long-term memory mechanisms~\cite{zhong2024memorybank}, continuous memory encoders~\cite{wu2025towards}, and cross-modal fusion modules~\cite{zhou2025learning}, show promise, they are designed for open-domain conversation and fall short in capturing the structured dependencies and persistent references typical of \gls{cps} modeling workflows.
%
To support chatbot-based modeling assistants, future research should therefore focus on developing memory architectures tailored to interactive modeling, capable of retaining long-term context, providing semantic grounding across multi-modal inputs (e.g., diagrams and textual descriptions), and tracking the evolution of \gls{cps} representations across modeling stages.

\textbf{Embedded Support in the Modeling Workflow.} 
\glspl{fm} can be embedded directly into \gls{cps} modeling environments to provide structured, real-time, and knowledge-aware assistance throughout the modeling lifecycle. Unlike traditional software systems, \gls{cps} design requires integrating domain-specific knowledge, such as physical dynamics, control theory, hardware constraints, and communication protocols, and managing inherently multimodal inputs, including natural language descriptions, architecture diagrams, simulation traces, and structured configuration data. 
As modelers edit diagrams, parameters, or behavior blocks, \glspl{fm} can observe these actions, interpret partial models, and offer context-aware suggestions grounded in engineering principles. 
These capabilities help ensure modeling decisions remain consistent, valid, and aligned with cross-domain requirements.
Beyond validating and refining models, \glspl{fm} can support partial automation of code generation tasks. 
\glspl{fm} can produce corresponding modeling constructs in SysML, Modelica, or similar languages and complete missing guards or transitions in behavioral models, infer likely parameter values, or auto-generate interface definitions based on previously defined components and their dependencies. This is especially useful when the modeler is not deeply familiar with a particular modeling language or domain formalism, reducing entry barriers and enabling broader participation.
%
\revision{To enable such embedded support, \glspl{fm} should be fine-tuned on \gls{cps}-specific multimodal artifacts and integrated with modeling tools to access model state and inject suggestions dynamically. Crucially, \gls{fm} outputs must be verified using simulation, constraint checking, or formal analysis to ensure semantic fidelity to physical and control constraints.}


\textbf{End-to-End Model Generation from Natural Language.} 
With increasing \gls{fm} capabilities, fully automated end-to-end model generation from natural language inputs is becoming feasible in certain structured domains.
Given a well-defined requirement specification, an \gls{fm} can autonomously identify relevant system components, infer their relationships, and organize them into coherent structural and behavioral representations. 
%
By interpreting such inputs holistically, \glspl{fm} could assist in rapidly constructing high-level models that reflect both the functional intent and the underlying physical constraints of the system. 
For example, a natural language description of a mobile robot tasked with navigating through dynamic environments could be translated into a structural model with sensor modules (e.g., LiDAR, cameras), motion control components, path planning logic, and obstacle avoidance behaviors.

Despite this potential, realizing robust and generalizable end-to-end CPS model generation remains challenging. \glspl{fm} must be fine-tuned on large, high-quality datasets that include multimodal and domain-specific artifacts, such as annotated requirement texts, formal specifications, model diagrams, and simulation logs, to better capture modeling semantics, domain conventions, and language syntax. 
Verification mechanisms are also critical. Without feedback from simulation, constraint checking, or structural validation, generated models may appear correct but fail under execution. 
In addition, techniques such as prompt engineering and human-in-the-loop refinement can improve generation robustness and adaptability to diverse modeling scenarios. 
For CPS in particular, future research should also focus on handling hybrid system representations (e.g., discrete and continuous models), ensuring physical realism, and aligning the generated models with both functional and physical domain constraints.

\subsubsection{\gls{fm}-based Mathematical Model Generation.}\label{subsubsec:fm4mathmodel} 

\begin{table}[htbp]
{\color{black}
\centering
\small
\caption{\revision{Actionable Research Opportunities Corresponding to the Identified Challenges for Foundation Models-based Mathematical Model Generation (Section~\ref{subsubsec:fm4mathmodel})}}
\label{tab:fm-math-model-opportunities}
\renewcommand{\arraystretch}{1.3} 
\begin{tabular}{|p{0.06\textwidth}|p{0.22\textwidth}|p{0.27\textwidth}|p{0.11\textwidth}|p{0.21\textwidth}|}
\hline
\rowcolor{black!70}
\color{white}\textbf{ID.} & \color{white}\textbf{Research Question} & \color{white}\textbf{Methodology} & \color{white}\textbf{Baseline(s)} & \color{white}\textbf{Evaluation Metrics} \\
\hline
\rowcolor{white}
$\mathrm{Ch}_{\mathrm{DM1}}$, $\mathrm{Ch}_{\mathrm{DM2}}$ & How can \gls{fm} capabilities be enhanced for multi-domain mathematical reasoning and symbolic-numerical \gls{cps} model generation? & Extend \glspl{fm} with mathematical reasoning modules and symbolic-numerical processing; train \glspl{fm} on multi-domain \gls{cps} mathematical models and simulation datasets for each specific domain. & General-purpose \glspl{fm} without fine-tuning, numerical solvers; & Accuracy of generated models, consistency with domain-specific constraints, efficiency, expert validation. \\
\rowcolor{gray!20}
$\mathrm{Ch}_{\mathrm{DM2}}$, $\mathrm{Ch}_{\mathrm{DM4}}$, $\mathrm{Ch}_{\mathrm{DM5}}$ & How can adaptive \gls{fm}-assisted workflows incrementally update mathematical models in response to evolving requirements and environmental conditions? & Implement adaptive pipelines in which \glspl{fm} updates model parameters and structure in response to requirement changes and environmental inputs, and integrate feedback loops for continuous learning. & Model regeneration from scratch, rule-based model update tools; & Model update accuracy, responsiveness to changes, reduction in manual intervention. \\
\hline
\end{tabular}
}
\end{table}

Mathematical modeling plays a critical role in CPS design, especially for capturing continuous dynamics, control rules, and system constraints. However, this task presents unique challenges for \glspl{fm}. First, \gls{cps} systems span multiple domains, e.g., mechanics, thermodynamics, and signal processing, each with its own modeling abstractions and mathematical formalisms~\cite{seshia2016design}. Second, the source data used for modeling can be highly heterogeneous, including time-series signals, physical measurements, structured design parameters, and textual specifications~\cite{amrani2021multi}. Interpreting and transforming this data into accurate mathematical expressions requires strong analytical, symbolic, and numerical reasoning skills. Finally, CPS models must adapt to changing system requirements or environmental conditions~\cite{muccini2016self}, placing additional demands on FMs to generate adaptive or parameterized models that remain valid over system variations.

FMs have shown preliminary capabilities in symbolic mathematics, numerical reasoning, and code generation, offering potential to assist in CPS mathematical modeling~\cite{imani2023mathprompter,ying2024internlm}. 
However, existing models still face limitations in reasoning across multiple physical domains, understanding diverse data types and formats, and adapting models to changing system conditions~\cite{guo2024controlagent,kevian2024capabilities}. 
To address these challenges, future work should focus on enhancing the mathematical reasoning capabilities of FMs, either by fine-tuning on domain-specific modeling tasks or by developing specialized FM variants with stronger analytical foundations. 
In addition, integrating external tools such as symbolic solvers, simulation engines, or hybrid neuro-symbolic reasoning modules may help bridge the gap between abstract model generation and executable system design. 
To handle dynamic changes in CPS environments, future work should explore how FMs can incrementally update mathematical models, e.g., adjusting equations or constraints, without regenerating them entirely, ensuring responsiveness and model consistency.
Moreover, realizing robust \gls{fm}-supported mathematical modeling will require new benchmarks, evaluation metrics, and co-designed workflows that reflect the complexity of real-world CPS scenarios.

\subsubsection{\glspl{fm} for \gls{cps} Digital Twin Creation.}\label{subsubsec:fm4dt}

\begin{table}[htbp]
{\color{black}
\centering
\small
\caption{\revision{Actionable Research Opportunities Corresponding to the Identified Challenges for \gls{cps} Digital Twin Creation with Foundation Models (Section~\ref{subsubsec:fm4dt})}}
\label{tab:fm-dt-opportunities}
\renewcommand{\arraystretch}{1.3} 
\begin{tabular}{|p{0.06\textwidth}|p{0.23\textwidth}|p{0.27\textwidth}|p{0.12\textwidth}|p{0.20\textwidth}|}
\hline
\rowcolor{black!70}
\color{white}\textbf{ID.} & \color{white}\textbf{Research Question} & \color{white}\textbf{Methodology} & \color{white}\textbf{Baseline(s)} & \color{white}\textbf{Evaluation Metrics} \\
\hline
\rowcolor{white}
$\mathrm{Ch}_{\mathrm{DM3}}$, $\mathrm{Ch}_{\mathrm{DM5}}$ & How can \glspl{fm} be fine-tuned to automatically generate high-fidelity \gls{dt} models and capabilities from \gls{cps} design specifications? & Train \glspl{fm} on paired datasets of \gls{cps} design specifications and validated digital twin models, generate models automatically with minimal manual intervention. & Manual \gls{dt} construction; & Fidelity of generated \gls{dt}, compliance with design specifications, reduction in manual effort. \\
\rowcolor{gray!20}
$\mathrm{Ch}_{\mathrm{DM4}}$, $\mathrm{Ch}_{\mathrm{DM5}}$ & How can \glspl{fm} be fine-tuned to act as live \glspl{dt} of \glspl{cps}? & Develop real-time \gls{fm} pipelines integrating sensor and operational data streams to simulate \gls{cps} behavior continuously. & Traditional \gls{dt} models, ML-based \gls{dt} models; & Accuracy of real-time predictions, \gls{fm} inference time, consistency with observed \gls{cps} behavior. \\
\rowcolor{white}
$\mathrm{Ch}_{\mathrm{DM1}}$, $\mathrm{Ch}_{\mathrm{DM3}}$ & How can cross-formalism \gls{dt} translation (e.g., SysML to/from OpenModelica) be enabled to ensure interoperability and reuse? & Implement \gls{fm}-based translation techniques between modeling formalisms, with semantic alignment and verification techniques. & Deep learning, rule-based methods; & Translation accuracy, preservation of model semantics. \\
\rowcolor{gray!20}
$\mathrm{Ch}_{\mathrm{DM3}}$, $\mathrm{Ch}_{\mathrm{DM5}}$ & How can human-in-the-loop verification and validation ensure trustworthy \glspl{dt}? & Integrate interactive \gls{fm} interfaces allowing experts to inspect, validate, and correct generated \glspl{dt} interactively, capture feedback for iterative model improvement. & Search-based, Deep learning-based, Other non-ML; & Expert validation scores, reduction in the number of required interventions, correctness, user satisfaction. \\
\rowcolor{white}
$\mathrm{Ch}_{\mathrm{DM1}}$, $\mathrm{Ch}_{\mathrm{DM5}}$ & How can \gls{fm}-based techniques identify, model, and quantify uncertainties in \gls{cps} \glspl{dt} using diverse multimodal data? & Fine-tune \glspl{fm} to analyze sensor, textual, and simulation data to detect and quantify uncertainties, incorporate uncertainty-aware modeling methods. & Manual, traditional uncertainty modeling methods; & Accuracy of uncertainty quantification, coverage of uncertain scenarios, robustness of \glspl{dt} predictions. \\
\rowcolor{gray!20}
$\mathrm{Ch}_{\mathrm{DM4}}$, $\mathrm{Ch}_{\mathrm{DM5}}$ & How can risk-aware, fine-tuned \glspl{fm} proactively integrate and manage uncertainty to maintain high-fidelity \glspl{dt}? & Develop \gls{fm} pipelines that monitor and adjust \gls{dt} models based on risk assessment and uncertainty propagation; simulate mitigation strategies. & rule-based risk assessment tools; & \gls{dt} fidelity under uncertainty, reduction in unexpected errors, risk coverage metrics. \\
\hline
\end{tabular}
}
\end{table}

Digital Twins (DTs) have become increasingly important in supporting the design of CPSs, offering simulation, analysis, and early validation capabilities before physical deployment~\cite{bachelor2019model,wang2021unified}. Lutze~\cite{lutze2020digital} highlights that DTs can enhance software design and lifecycle management by improving traceability, adaptability, and support for agile development compared to traditional V-model~\cite{johansson1999v} approaches. Besides, model-based DTs and digital threads can transform model-based design processes by improving traceability, interoperability, and tool integration~\cite{bachelor2019model}.
%
However, creating effective DTs during the CPS design phase remains a complex and resource-intensive task. 
First, constructing DT models often requires significant manual effort using modeling languages such as SysML or OpenModelica, demanding deep domain expertise and limiting scalability~\cite{wagner2023using,wilking2022sysml,chen2022enhancing}. 
Second, designing DT capabilities, such as predictive analytics or anomaly detection, typically depends on hand-crafted AI models, which lack automation and generalizability~\cite{mihai2022digital}. 
Finally, each CPS design usually necessitates a custom-tailored DT, resulting in high development cost and limited reuse~\cite{attaran2023digital,fett2023literature}. 
FMs, with their ability to generalize across tasks and domains, present a promising solution to reduce manual effort, enhance model reusability, and accelerate CPS design workflows. 
In the following, inspired by the work of Ali et al.~\cite{ali2024foundation}, we explore how FMs can support DT creation during CPS design: (1) assisting in the generation of DTs, and (2) fine-tuning FMs as DTs within the design loop.

\textbf{Generation of Digital Twins.}
During the design phase, FMs can assist in automatically generating both the DT models and their capabilities. 
For instance, language-based FMs can suggest or auto-complete structural elements of DT models in SysML or OpenModelica based on design documents or system specifications, thereby accelerating the creation of executable DT representations of CPS designs. 
They can also act as interactive assistants, guiding engineers not only in building simulation environments but in shaping DTs that faithfully mirror both the structure and behavior of the target system. 
On the DT capability side, FMs can help generate or configure AI components that enhance DT functionality, such as predictive simulation or early fault modeling, ensuring that the DT evolves beyond a static model into a dynamic DT representation of the CPS during design. 
In this way, FMs do not simply aid modeling tasks but directly contribute to reducing the manual effort required to construct high-fidelity DTs during CPS design.

Future work should focus on adapting FMs not only to modeling languages but also to the broader requirements of DT creation. This involves fine-tuning models on existing DT repositories, formal models, and CPS design specifications so that generated outputs align with both structural and behavioral aspects of DTs. Another promising direction is the automated translation between different modeling formalisms (e.g., SysML, OpenModelica, Simulink), which would support interoperability and enable DTs to integrate seamlessly across design environments. Beyond structural generation, research is needed on how FMs can support DT capability development, such as anomaly prediction, scenario realism assessment, and design-time optimization, thereby making DTs active assets in the design process rather than static models. Finally, human-in-the-loop approaches, where engineers iteratively refine FM-generated DTs, could ensure both fidelity and usability, positioning FMs as intelligent collaborators for creating high-quality DTs during CPS design.

\textbf{Fine-tuned \glspl{fm} as Digital Twins.}
An advanced approach is to fine-tune an FM so that it directly serves as a DT during CPS design. Instead of generating separate models and capabilities, the fine-tuned model itself can represent the system’s behavior, predict outcomes, and support reasoning tasks. 
Depending on the context, such a model may function as the DT model (capturing structure and behavior), as the DT capability (providing predictions, anomaly detection, or optimization), or as an integrated whole-DT solution. 
For example, a fine-tuned FM trained on design specifications and domain data could simulate design scenarios, anticipate potential failures, or recommend optimizations, thereby embedding intelligent decision support directly into the CPS design workflow. 
This approach significantly reduces the need for explicit, manually maintained DT models, while enabling DTs that continuously learn and adapt as design data evolves.

Future directions for this case should concentrate on systematically exploring how fine-tuned FMs can operate as design-time DTs with sufficient fidelity, efficiency, and interpretability. 
A key line of research is to establish guidelines for selecting suitable FMs for different CPS design contexts, whether text-based LLMs for processing specifications, vision-language models for analyzing simulation environments, or multimodal models for integrating heterogeneous design data. 
Another priority is ensuring that fine-tuned DTs can operate under design constraints, such as limited computation resources, real-time reasoning needs, and safety requirements, while still providing actionable feedback to engineers. 
Moreover, verification and validation techniques must be adapted to FM-based DTs, including methods to assess their fidelity in representing CPS behavior and their trustworthiness in safety-critical scenarios. 
Finally, hybrid approaches, where fine-tuned models complement but do not fully replace explicit DT representations, may provide a balanced path forward, combining the automation and adaptability of FMs with the transparency and control of traditional DT models to better support early-stage CPS design decisions.

\textbf{\glspl{fm} for dealing with uncertainty in DTs of \glspl{cps}.}
Uncertainty is an inherent characteristic of \glspl{cps} arising from various sources such as hardware, software, and environmental factors~\cite{asmat2023uncertainty}. 
A critical challenge in operating \gls{cps} DTs is maintaining their fidelity by precisely modeling and incorporating these uncertainties~\cite{sartaj2024uncertainty}. 
Addressing this challenge requires prior knowledge of various aspects, including sources of uncertainty, their impacts, and their probability distributions~\cite{mun2010modeling,zhang2016understanding}. 
However, a lack of comprehensive knowledge about uncertainties often limits the ability to operate high-fidelity DTs effectively~\cite{sartaj2024uncertainty}. 
With recent advances in identifying uncertainties with LLMs~\cite{sartaj2025identifying}, foundation models offer the potential to take this capability even further. 
Building on this, foundation models can be leveraged to develop advanced techniques that utilize diverse \gls{cps} data and artifacts to derive uncertainty sources, assess their impacts, and estimate probability distributions. 
Another promising direction is fine-tuning foundation models to develop risk models that enhance DTs by proactively integrating and managing uncertainties~\cite{ali2024foundation}. 
Moreover, uncertainty is intrinsic in foundation models due to their non-deterministic nature. 
For DTs developed using foundation models (e.g., foundation models fine-tuned as DTs~\cite{ali2024foundation}), it becomes essential to quantify and manage this uncertainty to ensure their fidelity. 
As highlighted by~\citet{ali2024foundation}, this represents a significant research direction, focusing on techniques to effectively assess and quantify uncertainty in foundation models when used in the creation of DTs.

\subsubsection{\glspl{fm} for CPS Model Execution Support.}\label{subsubsec:fm4modelexec} 


Executable models are essential for verifying CPS behavior, supporting simulation, control validation, and early-stage testing~\cite{andre2023formalizing}. Unlike traditional software execution, CPS model execution involves hybrid dynamics, real-time constraints, and multimodal inputs, requiring systems to react to sensor data, physical events, and control feedback~\cite{hehenberger2016design}. 
FMs have the potential to support this process by assisting both the execution of model logic and the generation of realistic input data, making it a promising direction for future exploration. In this section, we discuss two key aspects: (1) enabling FMs to support executable models via code generation and runtime evaluation, and (2) using FMs to synthesize realistic data to drive model execution.

\begin{table}[h!]
{\color{black}
\centering
\small
\caption{\revision{Actionable Research Opportunities Corresponding to the Identified Challenges for \gls{cps} Model Execution Support with Foundation Models (Section~\ref{subsubsec:fm4modelexec})}}
\label{tab:fm-model-execution-opportunities}
\renewcommand{\arraystretch}{1.3} 
\begin{tabular}{|p{0.06\textwidth}|p{0.20\textwidth}|p{0.28\textwidth}|p{0.12\textwidth}|p{0.20\textwidth}|}
\hline
\rowcolor{black!70}
\color{white}\textbf{ID.} & \color{white}\textbf{Research Question} & \color{white}\textbf{Methodology} & \color{white}\textbf{Baseline(s)} & \color{white}\textbf{Evaluation Metrics} \\
\hline
\rowcolor{white}
$\mathrm{Ch}_{\mathrm{DM2}}$, $\mathrm{Ch}_{\mathrm{DM5}}$ & How can \glspl{fm} generate executable \gls{cps} model code that satisfies domain-specific constraints and supports runtime checks? & Fine-tune FMs on \gls{cps} model and code template datasets; integrate constraint-checking and code-validation mechanisms. & Deep learning, natural language processing, other non-ML; & Correctness of generated code, compliance with constraints, runtime error detection, reduction in development effort. \\
\rowcolor{gray!20}
$\mathrm{Ch}_{\mathrm{DM4}}$, $\mathrm{Ch}_{\mathrm{DM5}}$ & How can \gls{fm}-based frameworks provide real-time assessment of executable \gls{cps} models during simulation and operation? & Develop interactive \gls{fm} pipelines to monitor model execution, detect deviations, and provide corrective feedback in real-time. & Traditional runtime monitors, search/ML-based validation; & Accuracy of anomalies detected, response time, improvement in simulation fidelity. \\
\rowcolor{white}
$\mathrm{Ch}_{\mathrm{DM1}}$, $\mathrm{Ch}_{\mathrm{DM2}}$ & How can multimodal \gls{fm}based methods generate realistic synthetic data to drive \gls{cps} model execution across diverse scenarios? & Fine-tune \glspl{fm} on multimodal datasets (text, sensor logs, diagrams, videos, audios) to synthesize data for simulation and testing. & Deep learning, search-based, reinforcement learning; & Realism and diversity of synthetic data, coverage of edge cases. \\
\rowcolor{gray!20}
$\mathrm{Ch}_{\mathrm{DM1}}$, $\mathrm{Ch}_{\mathrm{DM5}}$ & How can metrics and methods be developed to assess and improve the realism, diversity, and validity of \gls{fm}-generated execution data? & Define quantitative metrics for synthetic data quality; implement \gls{fm}-based evaluation and human-in-the-loop feedback loops to improve data generation quality iteratively. & Generic image or data quality metrics, manual expert evaluation; & Coverage of scenarios, realism score, diversity, expert validation score. \\
\hline
\end{tabular}
}
\end{table}

\textbf{Executable Model Code Generation and Runtime Evaluation.}
\Glspl{fm} can support the execution of executable models created for \glspl{cps}, such as UML state machines, by generating necessary low-level code and evaluating model constraints during runtime. Unlike high-level model construction, executable models must define fine-grained behavioral logic, including transition actions, guard conditions, event handling, and internal state updates. These models also need to account for domain-specific constraints, such as timing requirements, safety invariants, and control logic correctness.
To this end, \glspl{fm} have the potential to generate action code for transitions, complete missing guards, and identify violations of runtime constraints through behavioral reasoning. For example, given a UML state machine describing a mobile robot's motion controller, an FM could synthesize missing actions like obstacle avoidance or turning behavior, and evaluate whether speed limits or safe stopping conditions are respected during simulated execution.

Although some studies have made preliminary explorations into generating executable simulation code from FMs~\cite{ren2025simugen,xiang2025modigen}, this research area has not yet been sufficiently investigated. Supporting execution-time reasoning requires that \glspl{fm} not only understand the syntax and semantics of executable modeling languages, but also reason about temporal behaviors, control dependencies, and domain-specific rules. 
Beyond fine-tuning, future work must also consider tighter integration with executable modeling tools and mechanisms for maintaining long-range behavioral context during model execution. 
Moreover, model execution should be coupled with verification strategies, such as integrating FM support with existing simulation environments, runtime monitors, or formal checkers, to enable real-time validation of system behavior and ensure correctness and traceability.

\textbf{Realistic Execution Data Generation.}
Many CPS models rely on real-world data inputs to simulate behavior or validate design decisions. FMs, especially those with multimodal capabilities, have the potential to generate realistic synthetic data, such as LiDAR point clouds, camera images, or sensor signals, to drive model execution. 
This is particularly relevant in domains such as autonomous vehicles, where the model's response to complex sensor input is a core part of system validation. For instance, an FM could generate a video simulating a pedestrian crossing scenario, which can be processed by a CPS perception module and used to test the behavior of an executable system model.

However, generating high-fidelity and semantically valid data is a nontrivial task. Synthetic inputs must not only appear realistic but also cover a broad distribution of scenarios (e.g., lighting conditions, obstacle types, sensor noise) and must be consistent with the model’s expected context. Furthermore, systematic evaluation metrics for synthetic data quality, such as coverage, diversity, and realism, are still lacking, especially for CPS use cases. Future directions include guiding FM-based data generation using simulation feedback or real-world data distributions to improve realism and contextual fidelity, developing domain-specific metrics for evaluating the quality of generated data, and integrating human-in-the-loop feedback to guide data synthesis.

\section{Foundation Models for Software Development} \label{sec:SoftwareDev4CPS}
\revision{This section explores how \glspl{fm} can aid software developers improve their productivity. Section~\ref{sec:SD_sota} introduces the background and an overview of the state of the art in the use of \glspl{fm} for Software Development in the context of \glspl{cps}. Upon this foundation, Section~\ref{sec:SD_challegnes} identifies the key challenges and outlines how these challenges can be addressed through a set of relevant research questions.
}

\revision{
\subsection{Background and State of the Art}\label{sec:SD_sota}
\Glspl{fm} are known for their code generation capabilities~\cite{liu2023your}. According to Sam Altman, internal OpenAI studies report that \glspl{llm} can increase software engineering productivity by threefold~\cite{openai_sam_altman}. These capabilities have been continuously enhanced, particularly with the introduction of reasoning models such as OpenAI's o3-mini~\cite{openai_sam_altman} and DeepSeek-R1~\cite{guo2025deepseek}, which demonstrate improved performance in generating high-quality and logically consistent code.

Despite this rapid progress, most existing \gls{fm}-based approaches have been evaluated primarily in general-purpose software development contexts~\cite{fan2023large, hou2024large, jiang2024survey}. In contrast, \glspl{cps} exhibit distinct characteristics, including tight integration with physical processes, real-time constraints, safety-critical requirements, and usually rely on proprietary or highly specialized programming languages. These characteristics limit the direct applicability of state-of-the-art \glspl{fm} to \gls{cps} software development.
}

\subsection{Challenges and Research Opportunities}\label{sec:SD_challegnes}

\revision{While \glspl{fm} offer significant potential to enhance \gls{cps} software development, their effective adoption introduces several critical challenges. As summarized in Table~\ref{tab:sd-challenges}, we identify four core directions that must be addressed to enable the systematic integration of \glspl{fm} into CPS software engineering processes. Furthermore, Table~\ref{tab:fm-sd-opportunities} outlines how \glspl{fm}-based approaches can be leveraged to tackle these challenges, highlighting research opportunities associated with each direction.
}

\begin{table*}[h]
{\color{black}
\centering
\small
\caption{\revision{Challenges in \glspl{cps} Software Development}}
\label{tab:sd-challenges}
\rowcolors{2}{gray!20}{white}
\begin{tabular}{p{0.15\textwidth} p{0.8\textwidth}}
\rowcolor{black!70}
\textcolor{white}{\textbf{Challenge ID}} &
\textcolor{white}{\textbf{Challenge Description}} \\

$\mathrm{Ch}_{\mathrm{SD1}}$ &
Ensuring the qualification of FMs for their use in safety-critical contexts is challenging due to strict safety standards, regulations, and certification requirements, as well as lack of consideration of Generative AI techniques in these standards. \\

$\mathrm{Ch}_{\mathrm{SD2}}$ &
Ensuring accurate and reliable domain-specific code generation is difficult since \glspl{cps} languages are often specialized and proprietary. \\

$\mathrm{Ch}_{\mathrm{SD3}}$ &
Improving efficiency in cloud-constrained environments requires careful adaptation and optimization of \glspl{fm}, while protecting sensitive and confidential information. \\

$\mathrm{Ch}_{\mathrm{SD4}}$ &
Adapting \glspl{fm} to code-review tasks tailored to \glspl{cps} needs poses challenges for incorporating domain knowledge and enforcing compliance. \\

\end{tabular}
}
\end{table*}

\begin{table}[htbp]
{\color{black}
\centering
\small
\caption{\revision{Actionable Research Opportunities Corresponding to the Identified Challenges for \gls{cps} Software Development with Foundation Models}}
\label{tab:fm-sd-opportunities}
\renewcommand{\arraystretch}{1.3}
\begin{tabular}{|p{0.06\textwidth}|p{0.23\textwidth}|p{0.27\textwidth}|p{0.12\textwidth}|p{0.20\textwidth}|}
\hline
\rowcolor{black!70}
\color{white}\textbf{ID.} & \color{white}\textbf{Research Question} & \color{white}\textbf{Methodology} & \color{white}\textbf{Baseline(s)} & \color{white}\textbf{Evaluation Metrics} \\
\hline
\rowcolor{white}
$\mathrm{Ch}_{\mathrm{SD1}}$ & How can \glspl{fm} be fine-tuned to generate safety-critical \gls{cps} software that complies with real-time and operational constraints? & Fine-tune \glspl{fm} on domain-specific \gls{cps} codebases and integrate simulation-in-the-loop or runtime monitoring mechanisms to validate generated software under timing and safety constraints. & Manual implementation, rule-based synthesis; & Timing constraint violations, functional correctness, defect density, compliance with safety requirements. \\
\rowcolor{gray!20}
$\mathrm{Ch}_{\mathrm{SD2}}$ & How can \glspl{fm} support domain-specific code generation for heterogeneous \gls{cps} architectures (e.g., embedded systems, robotic middleware)? & Train \glspl{fm} on paired datasets of requirements/models and domain-specific implementations to generate domain-specific code. & Manual development, template-based code generation; & Code correctness, integration success rate, reduction in development time, and manual effort. \\
\rowcolor{white}
$\mathrm{Ch}_{\mathrm{SD3}}$ & How can lightweight or private \glspl{fm} be utilized to generate proprietary \gls{cps} software while preserving confidentiality? & Develop and deploy on-premise or compressed \gls{fm} variants fine-tuned on proprietary datasets for secure code generation in industrial environments. & Cloud-based \glspl{fm}, manual implementation; & Code quality, latency, data privacy compliance, resource utilization. \\
\rowcolor{gray!20}
$\mathrm{Ch}_{\mathrm{SD1}}$, $\mathrm{Ch}_{\mathrm{SD2}}$ & How can \glspl{fm} assist in verifying and validating generated control logic in \glspl{cps}? & Integrate \gls{fm}-generated control software with formal verification tools, simulation environments, or runtime monitoring frameworks to assess behavioral correctness. & Manual code review, testing-based validation; & Behavioral correctness, simulation success rate, number of detected violations or inconsistencies. \\
\rowcolor{white}
$\mathrm{Ch}_{\mathrm{SD4}}$ & How can \glspl{fm} maintain traceability between requirements, design models, and generated \gls{cps} software? & Utilize few-shot learning and traceability-aware prompting to automatically link generated code with upstream development artifacts. & Manual traceability analysis; & Precision/recall of traceability links, reduction in manual effort, traceability completeness. \\
\rowcolor{gray!20}
$\mathrm{Ch}_{\mathrm{SD4}}$ & How can \glspl{fm} support automated code review and documentation for certification-ready \gls{cps} software? & Apply \glspl{fm} to analyze generated source code, detect potential issues, and generate compliance-oriented documentation aligned with safety standards. & Manual code review; & Number of detected defects, documentation completeness, reviewer agreement rate. \\
\hline
\end{tabular}
}
\end{table}

\subsubsection{Safect-Critical Code Development}
Many \glspl{cps} are safety-critical, and thus must be certified, which also requires software development tools to be qualified by a regulatory body~\cite{conrad2010qualifying}. For instance, when compiling safety-critical code, the employed compiler should have been qualified by a functional safety regulatory agency. Different qualification categories exist~\cite{conrad2010qualifying}, and \gls{fm}-based code generation would require the most strict one as the code it generates might be buggy.

Further, each new version of the tool has to be re-qualified. All this introduces new challenges and aspects to consider when adopting \gls{fm}-based approaches. First, the tools need to carefully follow standards' rules; this can involve aspects like having redundancy for code generation, or checking that the \gls{fm} is robust to minor prompt changes. Second, standards need to be redefined and tailored to reconsider \gls{fm}-based techniques in developing safety-critical code; this needs interdisciplinary collaborations between regulators and AI experts to see how to adapt standards (e.g., forcing the use of uncertainty to monitor the FM's confidence when generating code). Lastly, while we see rapid advancements of \glspl{llm}, standards establish that each new version of the tool needs re-qualification; qualification processes are lengthy and expensive, which poses new challenges to \gls{fm} developers to make their new model versions re-qualified. Regulatory bodies should also research lighter processes to re-qualify fine-tuned LLM versions with enhanced code generation capabilities. 

\subsubsection{Domain Specific Code Generation}
Unlike mainstream general-purpose programming languages (e.g., Python, C), many \glspl{cps} (e.g., robots and programmable logic controllers) rely on their own proprietary coding language. These languages typically lack large, openly available code corpora, reducing the effectiveness of FMs trained predominantly on general-purpose code. A promising research direction is the automatic synthesis of domain-specific datasets, either by generating synthetic examples from formal language specifications or by mining industrial repositories under strict confidentiality constraints. Transfer learning techniques can further bridge this gap, for instance, by transferring representations learned from general-purpose programming languages to CPS-specific ones via adapter modules. Another research direction involves human-in-the-loop approaches, where FMs collaborate with engineers during code generation, creating curated datasets for future fine-tuning. This interplay between engineers and FMs can also help ensure adherence to domain standards, \revision{such as MISRA-C in automotive or DO-178C in avionics~\cite{Burden2016}.}

\subsubsection{CPS-specific Code Generation}
Much of \gls{cps} software, initially developed using simulation-based techniques, is later redeveloped near operation due to the reality gap. Examples include engineers reprogramming a robot, debugging CNC code along with the physical machine, or changing the control code of trains in new facilities. Cloud access is usually unavailable near the operation. Moreover, many companies hesitate to submit their code externally. We envision a direction for researching small language models of a few million parameters that can be directly deployed on the engineers' PCs and can still generate trustworthy \gls{cps} code. Lightweight adaptation methods such as quantization, pruning, or knowledge distillation can play a central role in enabling such deployment. Furthermore, continual on-device fine-tuning would allow these models to adapt to specific CPS environments without retraining from scratch. Running FMs locally also mitigates privacy risks by preventing sensitive code from leaving the organization’s infrastructure.

\subsubsection{Code Review}

Besides code generation, \glspl{fm} can serve as effective assistants for reviewing (safety-critical) software to be embedded on CPSs. Traditional static analysis tools are well-established, but they are usually limited to predefined rules and may overlook subtle issues arising from design choices, concurrency, or integration with hardware. In contrast, \glspl{fm} can complement such tools by providing semantic-level reasoning, flagging suspicious coding patterns, or suggesting alternative implementations that adhere to safety standards. An additional particularity of \gls{cps} development is its multidisciplinary nature: code is not only written by software engineers but also by electrical engineers, roboticists, control engineers, or mechanical engineers who may lack deep expertise in software engineering practices. In this context, \glspl{fm}-based reviewers can act as mediators, helping non-software experts align their code contributions with safety and coding standards.

A promising direction is hybrid review pipelines, where \glspl{fm} are combined with rule-based analyzers: while analyzers ensure compliance with standards such as MISRA-C or IEC 61508, \glspl{fm} can highlight higher-level design flaws or undocumented assumptions (e.g., identifying architectural anti-patterns such as bypassing safety wrappers to access actuators). Another important research path is explainability. For FM-based code review to be trusted in safety-critical \gls{cps}, engineers must understand why the model flags certain lines of code, which requires human-readable rationales or links to known standards violations. Finally, given the safety-critical nature of CPSs, their projects usually demand traceability. Consequently, \glspl{fm} should not only provide feedback but also generate structured reports that can be archived as part of the certification evidence.

\section{Foundation Models for \gls{cps} Testing} \label{sec:CPSTest}
\revision{This section explores how FMs can advance \gls{cps} testing. Section~\ref{subsec:test_stateart} introduces the background and the state of the art in the use of FMs for \gls{cps} software testing, while Section~\ref{subsec:testing_chall} identifies critical challenges in testing \glspl{cps} and outlines research opportunities and research questions to address them.
%
}





\revision{
\subsection{Background and State of the Art}\label{subsec:test_stateart}
CPS integrates hardware and software components that directly interact with the physical environment and is usually designed for safety-critical applications, such as transportation, healthcare, and industrial control. Thus, rigorous testing of CPS is crucial to ensure safe and reliable operation under various conditions, and simulation-based testing is a typical way that provides a controlled environment to evaluate system behavior under various conditions. Traditional simulation-based testing approaches employ techniques such as meta-heuristic search and reinforcement learning to generate critical conditions to test CPS~\cite{8722847,7050320}. However, such approaches show limited effectiveness when it comes to bridging the reality gap~\cite{birchler2024roadmap}, managing computational resources, and ensuring the explainability of the detected failures~\cite{gerking2019explainability}. 

\Glspl{fm} have been applied to support \gls{cps} software testing across several domains, including test generation for \glspl{av}~\cite{10529537,10.1145/3691620.3695037,lu2024realistic,lu2024multimodal,deng2023target}, testing of decision-making policies in robotics~\cite{xu2024exploring,10.1145/3663529.3663779}, and industrial \gls{cps}~\cite{lu2024assessing,10.1109/TSE.2024.3368208}. \Glspl{llm} have demonstrated capabilities in generating interpretable test scenarios and extracting relevant information from unstructured sources. Besides, \glspl{vlm} offer enhanced capabilities for synthesizing multimodal test inputs that combine visual context with semantic information~\cite{lu2024assessing,wu2024reality}. Despite these advances, applying \glspl{fm} to \gls{cps} testing introduces several notable challenges, which, together with key research opportunities, are discussed below.

}




\begin{table*}[t]
{\color{black}
\centering
\small
\caption{\revision{Challenges in Testing of \glspl{cps}}}
\label{tab:testing-challenges}
\rowcolors{2}{gray!20}{white}
\begin{tabular}{p{0.15\textwidth} p{0.8\textwidth}}
\rowcolor{black!70}
\textcolor{white}{\textbf{Challenge ID}} &
\textcolor{white}{\textbf{Challenge Description}} \\

$\mathrm{Ch}_{\mathrm{T1}}$ &
Ensuring test effectiveness and efficiency in dynamic \glspl{cps} environments is non-trivial due to continuous evolution and uncertain environments. \\

$\mathrm{Ch}_{\mathrm{T2}}$ &
Addressing inefficiencies in CPS software regression testing is resource-intensive due to the heterogeneous nature of CPSs, involving evolution of software (including AI components) and hardware updates. \\

$\mathrm{Ch}_{\mathrm{T3}}$ &
Absence of precise test oracles in CPS testing makes test automation difficult. \\

$\mathrm{Ch}_{\mathrm{T4}}$ &
Automating the transformation from test case specifications to executable tests is challenging due to the heterogeneous set of tools and formalisms used across different CPS domains. \\

$\mathrm{Ch}_{\mathrm{T5}}$ &
Bridging the reality gap between simulation and real-world behavior is problematic since simulation cannot fully capture the physical environment and its inherent uncertainties. \\

$\mathrm{Ch}_{\mathrm{T6}}$ &
Handling the complexity of testing interactive control systems is demanding due to constraints such as real-time operation and feedback loops. \\

\end{tabular}
}
\end{table*}






\subsection{Challenges and Research Opportunities}\label{subsec:testing_chall}
\revision{While \glspl{fm} have shown potential in advancing \gls{cps} testing, their effective adoption introduces several critical challenges. As summarized in Table~\ref{tab:testing-challenges}, we identify six core challenges that must be addressed to enable the systematic integration of \glspl{fm} into \gls{cps} testing processes. For each challenge, we outline research opportunities, research questions, and potential methodologies, as well as baselines and metrics for assessing the cost-effectiveness of these methodologies, which are presented in Tables~\ref{tab:fm-testing-opportunities},~\ref{tab:fm-testing-selection},~\ref{tab:fm-testing-oracle},~\ref{tab:fm-testing-reality}, and~\ref{tab:fm-testing-interactive}.}

\subsubsection{FM-based Test Generation and Test Support}\label{subsubsec:testsupport}

\begin{table}[htbp]
{\color{black}
\centering
\small
\caption{\revision{Actionable Research Opportunities Corresponding to the Identified Challenges for \gls{cps} Testing with Foundation Models: Test Generation and Test Support (Section~\ref{subsubsec:testsupport})}}
\label{tab:fm-testing-opportunities}
\renewcommand{\arraystretch}{1.3}
\begin{tabular}{|p{0.06\textwidth}|p{0.23\textwidth}|p{0.27\textwidth}|p{0.12\textwidth}|p{0.20\textwidth}|}
\hline
\rowcolor{black!70}
\color{white}\textbf{ID.} & \color{white}\textbf{Research Question} & \color{white}\textbf{Methodology} & \color{white}\textbf{Baseline(s)} & \color{white}\textbf{Evaluation Metrics} \\
\hline
\rowcolor{white}
$\mathrm{Ch}_{\mathrm{T1}}$ & How can \glspl{fm} automatically generate high-fidelity \gls{cps} test scenarios and realistic test data under diverse environmental conditions? & Fine-tune \glspl{fm} on historical \gls{cps} test data and environmental data to generate diverse test scenarios. & Search-based, ML-based test generation. & Test scenario coverage, realism of generated tests, and fault detection rate. \\
\rowcolor{gray!20}
$\mathrm{Ch}_{\mathrm{T1}}$, $\mathrm{Ch}_{\mathrm{T6}}$ & How can hybrid approaches combining \glspl{fm} with traditional methods improve \gls{cps} test generation? & Integrate \glspl{fm} with techniques such as search-based and model-based testing to enhance scenario diversity, fault detection, coverage, as well as incorporating knowledge from expert models. & Search-based testing alone, model-based testing alone. & Fault detection rate, scenario diversity, and reduction in manual test efforts. \\
\rowcolor{white}
$\mathrm{Ch}_{\mathrm{T4}}$ & How can \glspl{fm} transform test case specifications into executable tests efficiently? & Develop \gls{fm}-based solutions to understand test specifications and generate executable test scripts for a given \gls{cps} testing framework. & Natural language processing, manual transformation. & Correctness of executable tests, reduction in manual coding, and execution success rate. \\
\hline
\end{tabular}
}
\end{table}

\textbf{FM-based Test Generation and Test Data Augmentation}.
\gls{cps} test generation is time-consuming and resource-intensive, particularly due to the complexity and interdependencies inherent in \glspl{cps}. Zhou et al.~\cite{zhou2018review} investigated and identified key challenges for testing CPS. Despite the rapid development of advanced testing methods, bottlenecks remain, especially for complex CPSs requiring reliability, security, and resilience. Key challenges include managing large and complex state spaces, addressing state space explosion, handling uncertainty during testing, ensuring real-time performance, and designing effective test oracles.
\revision{\glspl{fm} offer a promising solution by automating and enhancing test generation~\cite{schafer2023empirical,wang2024software}.} They can generate diverse, high-fidelity test scenarios that cover edge cases, simulate realistic sensor inputs and environmental conditions, and model system interactions, addressing state space and uncertainty challenges. \glspl{fm} can also adapt test sequences in real time to meet timing constraints and assist in designing test oracles by predicting expected system outputs based on learned behaviors. For example, \glspl{fm} can generate a diverse set of high-fidelity test scenarios that cover corner cases and rare events, helping to manage large and complex state spaces and mitigating state space explosion. By leveraging their ability to model multimodal data, \glspl{fm} can simulate realistic sensor inputs, environmental conditions, and system interactions, which allows for comprehensive uncertainty modeling. In addition, \glspl{fm} can be used to dynamically generate and adapt test sequences in real time, supporting testing under strict timing constraints and ensuring that system responses are evaluated within operational deadlines. 
\revision{Besides, \glspl{fm} can enrich the test suite by performing noising or denoising on test data~\cite{wang2025llm4dsr,zhou2024can}.} They can simulate realistic perturbations, such as sensor inaccuracies or environmental disturbances, to create challenging noisy data. While \glspl{fm} can also reconstruct clean signals from corrupted inputs, filtering out irrelevant noise while preserving critical features. This dual capability would allow generating challenging (noisy) and clean (noise-free) test data for \gls{cps}, thereby broadening the spectrum of conditions under which the \gls{cps} is evaluated. \looseness=-1

\textbf{Integrate FMs with Traditional Testing Techniques}.
Ahn et al.~\cite{ahn2024large} note that FMs exhibit limited performance in numerical problem-solving, often struggling with accuracy, consistency, and complex calculations. When it comes to CPS testing, numerical reasoning is critical for generating concrete configuration parameters from logical scenarios defined by parameter ranges.  Concrete scenarios with specific parameter values from logical scenarios defined by parameters and parameter ranges are crucial for the precise simulation of CPS dynamics and its environment. 
This limitation is critical because precise simulation of \gls{cps} dynamics and their environment relies on accurate numerical instantiation.
Consequently, integrating traditional techniques—such as search-based methods and numerical solvers—with \glspl{fm} becomes necessary to provide a comprehensive solution. These techniques can complement the capabilities of \glspl{fm} by enabling precise numerical computation, enhancing the fidelity of \gls{cps} simulations. For example, integrating \glspl{fm} with search-based algorithms offers a promising hybrid approach for CPS testing. \glspl{fm} can generate high-level logical scenarios from requirements and provide heuristics to guide test generation, while search-based algorithms systematically explore parameter spaces to produce concrete configurations and ensure coverage of edge cases. 
This combination leverages the reasoning and generative capabilities of \glspl{fm} while ensuring numerical accuracy and systematic exploration, enhancing both the fidelity and effectiveness of CPS testing.
Such a hybrid approach would effectively leverage logical reasoning and mathematical computing abilities, addressing the limitations of \glspl{fm} in numerical problem-solving. However, there is a need for more research to build such hybrid approaches. 

\textbf{Transform Test Case Specifications to Test with FMs}.
Test case specifications (TCSs) play a critical role in \gls{cps} software testing~\cite{juhnke2021challenges,wang2015automatic}, as they provide structured descriptions of inputs, expected behaviors, and environmental conditions. TCSs are often written in natural language; thus, manually transforming TCSs into executable test cases is time-consuming, error-prone, and requires significant coding effort and domain knowledge. Moreover, manual approaches fail to capture the diversity required to validate complex systems in dynamic environments. To address this, automated transformation from TCSs to test cases has been studied. For example, Tao et al.~\cite{yue2015rtcm} proposed RTCM, a restricted test case modeling language with templates, rules, and keywords, along with a tool that generates either manual or executable test cases from structured TCSs. To support the runtime \gls{cps} test, Shi et al.~\cite{shi2021restricted} propose LiveTCM, which interacts with the \gls{cps} and its environment to generate and execute TCSs efficiently. 
\Glspl{fm} offer promising opportunities for further automating the transformation process by interpreting natural-language TCSs, reasoning about multimodal TCSs, and generating diverse executable test cases with minimal human intervention. \glspl{fm} can also adapt to evolving system behaviors and environmental conditions, potentially improving coverage, fault detection, and efficiency in CPS testing. 
For example, \glspl{fm} can translate high-level TCSs of environmental conditions, control signals, and system properties into executable scripts suitable for runtime verification. In addition, \glspl{fm} can support diversity and scalability in test generation by synthesizing variations of test cases from the same TCS. This includes altering environmental parameters, introducing disturbances, or modifying mission goals, thereby producing a richer set of test scenarios to assess CPS robustness under dynamic and uncertain conditions. By leveraging fine-tuning or domain adaptation, \glspl{fm} can also learn from historical test data and execution outcomes, continuously refining their ability to generate effective test cases over time.


\subsubsection{FM-based Test Selection, Minimization, and Prioritization}\label{subsubsec:test_select}
Regression testing for \gls{cps} software updates helps detect bugs in the modified version of the software. However, as test suites expand, redundant and overlapping test cases can lead to inefficiencies and increased execution times~\cite{menghi2020approximation, valle2023automated, abdessalem2020automated}. This issue is particularly acute for CPS, where software interacts continuously with dynamic, stochastic physical environments, making exhaustive testing impractical. Therefore, it is crucial to adopt approaches that can efficiently manage and optimize regression test suites. While traditional approaches demonstrate good performance~\cite{lu2021search,yoo2012regression,mukherjee2021survey} in test selection, minimization, and prioritization, \glspl{fm} offer a promising way forward by automatically analyzing the characteristics to 1) identify and prioritize the subset of tests most likely to detect failures early in the development cycle, 2) minimize the overall suite while preserving fault-detection effectiveness, and 3) optimize test execution order to maximize early fault detection and overall testing efficiency. For instance, \glspl{fm} can analyze patterns across historical test executions, system behaviors, and environmental conditions to predict which tests are most likely to detect failures, enabling smarter prioritization that accelerates feedback to developers. They can also reason over multimodal test data, such as sensor readings, control signals, and simulation traces, allowing for more accurate identification of redundant or overlapping tests and more effective test suite minimization. By learning from past executions, \glspl{fm} can adapt test selection, minimization, and prioritization strategies to evolving CPS behaviors and complex operating environments.

\begin{table}[htbp]
{\color{black}
\centering
\small
\caption{\revision{Actionable Research Opportunities Corresponding to the Identified Challenges for \gls{cps} Testing with Foundation Models: Test Selection, Minimization, and Prioritization (Section~\ref{subsubsec:test_select})}}
\label{tab:fm-testing-selection}
\renewcommand{\arraystretch}{1.3}
\begin{tabular}{|p{0.06\textwidth}|p{0.23\textwidth}|p{0.27\textwidth}|p{0.12\textwidth}|p{0.20\textwidth}|}
\hline
\rowcolor{black!70}
\color{white}\textbf{ID.} & \color{white}\textbf{Research Question} & \color{white}\textbf{Methodology} & \color{white}\textbf{Baseline(s)} & \color{white}\textbf{Evaluation Metrics} \\
\hline
\rowcolor{white}
$\mathrm{Ch}_{\mathrm{T2}}$ & How can \glspl{fm} prioritize tests based on provided cost and effectiveness criteria? & Fine-tune \glspl{fm} on historical test execution data, e.g., test execution time, failure patterns, fault detection rates, to enable them to prioritize test cases for execution. & Search-based testing, reinforcement learning. & Early fault detection rate, testing efficiency. \\
\rowcolor{gray!20}
$\mathrm{Ch}_{\mathrm{T2}}$ & How can \glspl{fm} minimize test suites while preserving effectiveness? & Implement \gls{fm}-based minimization algorithms that select representative tests while given effectiveness criteria such as coverage and fault detection, while also analyzing code, logs, and other test artifacts available. & Search-based, reinforcement learning. & Fault detection rate, reduction in test suite size, reduction in test execution time. \\
\hline
\end{tabular}
}
\end{table}

\subsubsection{FM-based Test Oracle Support}
The lack of test oracles or the absence of precise test oracles is common when testing \glspl{cps}~\cite{ayerdi2022performance,ayerdi2021generating}, as in many other software systems~\cite{isaku2024llms}. To this end, research is needed to explore the use of \glspl{fm} for generating test oracles, based on the assumption that they have learned from a wide range of information and can assist testers in devising test oracles for \gls{cps} testing. Moreover, to further support test oracles in the context of differential testing, one of the key methods for addressing the test oracle problem, \glspl{fm} can be customized to serve as a surrogate implementation of a \gls{cps}. This is an open research area that requires further investigation.

\begin{table}[htbp]
{\color{black}
\centering
\small
\caption{\revision{(c) Actionable Research Opportunities Corresponding to the Identified Challenges for \gls{cps} Testing with Foundation Models: Test Oracle Support}}
\label{tab:fm-testing-oracle}
\renewcommand{\arraystretch}{1.3}
\begin{tabular}{|p{0.06\textwidth}|p{0.23\textwidth}|p{0.27\textwidth}|p{0.12\textwidth}|p{0.20\textwidth}|}
\hline
\rowcolor{black!70}
\color{white}\textbf{ID.} & \color{white}\textbf{Research Question} & \color{white}\textbf{Methodology} & \color{white}\textbf{Baseline(s)} & \color{white}\textbf{Evaluation Metrics} \\
\hline
\rowcolor{white}
$\mathrm{Ch}_{\mathrm{T3}}$ & How can \glspl{fm} be used to generate \gls{cps} test oracles using knowledge from diverse sources? & Train \glspl{fm} on formal specifications, logs, and documentation to suggest expected outputs and detect deviations. & Manual oracle design. & Test oracle correctness and precision, fault detection, and reduction in manual oracle design. \\
\rowcolor{gray!20}
$\mathrm{Ch}_{\mathrm{T3}}$ & How can \glspl{fm} act as surrogate \gls{cps} implementations for differential testing to enhance automation, accuracy, and coverage? & Fine-tune \glspl{fm} to simulate \gls{cps} behavior as test oracles; compare outputs against multiple fine-tune \glspl{fm} as well as simulation models and implementations for automated validation. & Manual differential testing, simulation models, reference implementations. & Correctness and precision of surrogate oracles, coverage of scenarios, and improvement in fault detection. \\
\hline
\end{tabular}
}
\end{table}

\subsubsection{Evaluate and Bridge Reality Gap with FMs}
Testing \gls{cps} usually involves a simulated testing environment or requires synthesizing test inputs (e.g., images and videos), but ensuring the test scenarios or synthetic test inputs accurately represent all facets of the physical systems and environments is still an open problem. Salvato et al.~\cite{salvato2021crossing} surveyed reinforcement learning for robot control, focusing on the reality gap problem that hinders sim-to-real transfer. They reviewed existing approaches to mitigate reality gaps, discussed their limitations, and outlined promising directions for future research. In the AV domain, Stocco et al.~\cite{9869302} investigated the sim-to-real gap in AV testing, showing that simulation-based testing can overlook critical reality issues, while also identifying situations where physical testing is unnecessary because simulation results transfer reliably. \glspl{fm} offer promising opportunities to evaluate and bridge this gap. 
\revision{By leveraging their ability to model multimodal data, \glspl{fm} can assess the realism and fidelity of test scenarios and quantify how closely simulated conditions align with real-world environments~\cite{wu2024reality}.} They can do this by comparing visual, sensory, and contextual features, as well as by checking higher-level patterns such as consistent behavior, realistic environmental dynamics, and rare corner cases. This enables a more thorough evaluation of whether simulated inputs capture the complexity, variability, and uncertainty of real systems, helping to pinpoint gaps in test coverage and refine simulation models for greater accuracy.
Furthermore, by fine-tuning \glspl{fm} with domain-specific naturalistic data and environmental conditions, specialized \glspl{fm} can be developed to generate synthetic inputs that more faithfully represent real-world scenarios. These models can capture complex interactions among system components, environmental variations, and rare or extreme events, enabling the creation of diverse and realistic test cases. Such \gls{fm}-generated inputs not only improve the coverage and representativeness of testing but also reduce reliance on costly or risky physical trials, supporting safer and more efficient evaluation of CPS software.

\begin{table}[htbp]
{\color{black}
\centering
\small
\caption{\revision{(d) Actionable Research Opportunities Corresponding to the Identified Challenges for \gls{cps} Testing with Foundation Models: Evaluating and Bridging the Reality Gap}}
\label{tab:fm-testing-reality}
\renewcommand{\arraystretch}{1.3}
\begin{tabular}{|p{0.06\textwidth}|p{0.23\textwidth}|p{0.27\textwidth}|p{0.12\textwidth}|p{0.20\textwidth}|}
\hline
\rowcolor{black!70}
\color{white}\textbf{ID.} & \color{white}\textbf{Research Question} & \color{white}\textbf{Methodology} & \color{white}\textbf{Baseline(s)} & \color{white}\textbf{Evaluation Metrics} \\
\hline
\rowcolor{white}
$\mathrm{Ch}_{\mathrm{T5}}$ & How can multimodal \glspl{fm} quantify and improve the fidelity of simulated \gls{cps} test scenarios? & Train \glspl{fm} on multimodal simulation and real-world datasets based on carefully chosen metrics to compare simulation fidelity and suggest scenario refinements. & Manual fidelity assessment, domain randomization methods. & Improvement in simulation fidelity, scenario coverage. \\
\rowcolor{gray!20}
$\mathrm{Ch}_{\mathrm{T1}}$, $\mathrm{Ch}_{\mathrm{T5}}$ & How can \glspl{fm} generate realistic synthetic inputs capturing rare events and complex system-environment interactions? & Fine-tune \glspl{fm} on historical \gls{cps} logs, environmental data, and uncertain scenarios, and generate synthetic inputs for testing. & Search-based, reinforcement learning, manual. & Realism of generated inputs, improvement in fault detection. \\
\rowcolor{white}
$\mathrm{Ch}_{\mathrm{T5}}$ & How can \glspl{fm} predict simulation-to-real-world transferability to prioritize cases for physical testing? & Develop \gls{fm}-based predictive models mapping simulation scenarios to expected real-world outcomes to prioritize scenarios for physical testing. & Deep learning transferability predictors, manual expert selection. & Prediction accuracy, physical testing efficiency, reduction in cost. \\
\hline
\end{tabular}
}
\end{table}

\subsubsection{FMs for Testing Interactive Control Systems}
Interactive control systems, often referred to as human-machine interfaces, graphically display real-time data from the physical components of \gls{cps} to enable operators to monitor, interact with, and control these systems effectively. 
In safety-critical domains, such as unmanned aerial vehicles, testing these systems becomes essential to support informed decision-making and ensure the safety and reliability of \gls{cps} operations. 
The existing techniques in the literature are based on traditional image processing methods, such as computer vision and optical character recognition~\cite{sartaj2020cdst,sartaj2021testing}. 
These approaches often require significant manual effort in generating test scripts, executing simulations, processing image data to achieve high prediction accuracy, and evaluating test verdicts~\cite{sartaj2021testing}. 
Another significant challenge lies in adapting these approaches to new or evolved display systems with updated controls, which typically requires considerable tailoring to existing methods or the development of entirely new solutions from scratch~\cite{sartaj2024automated}. 
The advent of foundation models presents exciting opportunities to overcome these challenges. 
\revision{One promising direction is to develop LLM-based techniques to automatically generate test scenarios and executable scripts to test interactive control systems, such as~\cite{rosenbach2025automated}. 
Another direction involves developing VLM-based techniques to analyze test verdicts by comparing images and facilitating the evaluation process~\cite{zhang2024vision}.} 
Furthermore, a direction worth exploring is leveraging multimodal foundation models to generate diverse data, such as images and videos, facilitating the testing of interactive display systems without relying on computationally expensive simulations.

\begin{table}[htbp]
{\color{black}
\centering
\small
\caption{\revision{(e) Actionable Research Opportunities Corresponding to the Identified Challenges for \gls{cps} Testing with Foundation Models: Testing Interactive Control Systems}}
\label{tab:fm-testing-interactive}
\renewcommand{\arraystretch}{1.3}
\begin{tabular}{|p{0.06\textwidth}|p{0.23\textwidth}|p{0.27\textwidth}|p{0.12\textwidth}|p{0.20\textwidth}|}
\hline
\rowcolor{black!70}
\color{white}\textbf{ID.} & \color{white}\textbf{Research Question} & \color{white}\textbf{Methodology} & \color{white}\textbf{Baseline(s)} & \color{white}\textbf{Evaluation Metrics} \\
\hline
\rowcolor{white}
$\mathrm{Ch}_{\mathrm{T4}}$, $\mathrm{Ch}_{\mathrm{T6}}$ & How can \glspl{llm} automatically generate test scenarios and executable scripts for interactive control systems in \gls{cps}? & Fine-tune \glspl{llm} on control system specifications, prior test cases, test logs (if available), generate executable test scripts automatically for a given testing framework. & Search-based, reinforcement learning. & Test scenario coverage, correctness of scripts, reduction in manual effort. \\
\rowcolor{gray!20}
$\mathrm{Ch}_{\mathrm{T3}}$, $\mathrm{Ch}_{\mathrm{T6}}$ & How can \glspl{vlm} analyze and evaluate test verdicts from \gls{cps} visual systems for automated test assessment? & Fine-tune \glspl{vlm} on visual test logs and expected outcomes; automate verdict generation and evaluation. & Traditional image processing, manual assessment. & Correctness of automated verdicts, reduction in manual effort, detection of visual anomalies. \\
\hline
\end{tabular}
}
\end{table}

\section{Foundation Models for Debugging and Repair} \label{sec:CPSrepasir}
\revision{This section examines how \glspl{fm} can support developers of \glspl{cps} in debugging and repairing such systems. Section~\ref{sec:DR_back} first provides background and an overview of the state of the art in the use of \glspl{fm} for debugging and repair in the context of \glspl{cps}. Building on this foundation, Section~\ref{sec:DR_challenges} identifies the key challenges and outlines how they can be addressed through a set of relevant research questions.}

\revision{
\subsection{Background and State of the Art}\label{sec:DR_back}
\glspl{fm}, and in particular \glspl{llm}, have recently emerged as powerful enablers for automated debugging~\cite{10.1145/3293882.3330574,qin2025agentflscalingllmbasedfault,Wang_2024,Kang_2024} and repair~\cite{Wei_2023,10.1145/3180155.3180233,10.1109/ICSE48619.2023.00129,Xia_2022,Xia_2024,zhang2023gammarevisitingtemplatebasedautomated,islam2024mapcodermultiagentcodegeneration} due to their ability to learn from vast and heterogeneous datasets, including source code, execution logs, documentation, and other artifacts, as well as their multimodal capabilities. These models benefit from strong code understanding, language adaptability, and the ability to generate candidate patches based on learned patterns. A growing body of work explores \gls{llm}-based agents that unify fault localization and repair into a single workflow~\cite{zhong2024debuglikehumanlarge,lee2024unifieddebuggingapproachllmbased}. Such agents enhance traditional debugging approaches by leveraging \glspl{llm} to interpret error messages, analyze execution traces, and coordinate repair actions with greater flexibility. Their ability to understand and integrate outputs from both fault localization and repair stages represents a significant advancement over conventional, modular approaches.

In the context of \glspl{cps}, the use of \glspl{fm} has only recently begun to emerge. Early work such as FixDrive~\cite{sun2025fixdriveautomaticallyrepairingautonomous} illustrates the potential of learning-based techniques for repairing \gls{cps}-related software, particularly in autonomous driving systems. Nevertheless, the broader application of \glspl{fm} to \gls{cps} debugging and repair remains largely unexplored. As highlighted by Valle et al.~\cite{valle2025automated}, achieving fully automated \gls{cps} debugging and repair requires addressing several open challenges related to fault localization, scalability, testing cost, and repair guidance.
}

\subsection{Challenges and Research Opportunities}\label{sec:DR_challenges}

While \glspl{fm} have demonstrated strong potential in traditional software domains, \gls{cps}  debugging and repair introduce additional complexity from the interaction of software, hardware, and physical environments. These characteristics create significant challenges, presented in Table~\ref{tab:debug-repair-challenges}, but also open new research opportunities, as can be seen in Table~\ref{tab:fm-debugging-repair-opportunities}. In this section, we discuss four major challenges and outline how \glspl{fm} can contribute to addressing them.

\begin{table*}[h]
{\color{black}
\centering
\small
\caption{\revision{Challenges in \gls{cps} Debugging and Repair}}
\label{tab:debug-repair-challenges}
\rowcolors{2}{gray!20}{white}
\begin{tabular}{p{0.15\textwidth} p{0.8\textwidth}}
\rowcolor{black!70}
\textcolor{white}{\textbf{Challenge ID}} &
\textcolor{white}{\textbf{Challenge Description}} \\

$\mathrm{Ch}_{\mathrm{DR1}}$ &
Difficulties in effective fault localization due to complicated interactions between different software, hardware, and physical environments. \\

$\mathrm{Ch}_{\mathrm{DR2}}$ &
Handling a wide range of bugs is time- and resource-consuming due to possible failures emerging from software, AI components, and environmental uncertainties. \\

$\mathrm{Ch}_{\mathrm{DR3}}$ &
Expensive test executions, as reproducing faults may require large-scale simulations and physical experiments with real hardware. \\

$\mathrm{Ch}_{\mathrm{DR4}}$ &
Improving the fitness functions that guide repair algorithms is challenging because it requires accounting for multi-dimensional aspects such as cost, effectiveness, and safety across different CPS components. \\

\end{tabular}
}
\end{table*}

\subsubsection{Effective Fault Localization} 
Fault localization in \glspl{cps} is particularly difficult because faults often arise from interactions between software components and physical processes rather than isolated code defects. Moreover, CPS testing is commonly performed at the system level, which limits the applicability of traditional fault localization techniques. This challenge is further exacerbated by long-running tests and the prevalence of flaky simulators, which can distort test results and undermine localization accuracy~\cite{amini2024evaluating}. \glspl{fm} offer promising opportunities to address these issues by enabling multi-modal reasoning across software and physical layers. Recent work such as SemCoDer~\cite{ding2024semcoder} demonstrated that \glspl{llm} can be trained to perform deep semantic reasoning over source code. When combined with time-series or multi-modal models~\cite{Liang_2024}, \glspl{fm} could correlate failing code with sensor readings or system states observed prior to a failure. For example, a model could jointly analyze source code, execution logs, and sensor outputs to identify suspicious code.

Additionally, \glspl{llm} have shown strong capabilities in interpreting error logs and execution traces for software debugging~\cite{rafi2025multiagentapproachfaultlocalization}. Extending these techniques to CPS-specific errors requires training or fine-tuning models on CPS tracing datasets that include code, textual logs, and time-series data. A recent survey on this~\cite{Liang_2024} suggests the current \glspl{fm} can spot when a sensor starts failing and link that issue with the source code. Another promising direction is the integration of domain knowledge, such as physical constraints or safety requirements, into \glspl{fm}. Prior work indicates that \glspl{llm} can incorporate semantic rules during reasoning~\cite{ding2024semcoder}, which could help enforce known safety properties during fault localization. Finally, combining \glspl{fm} with traditional fault localization techniques may further improve effectiveness. For instance, Valle et al.~\cite{valle2025automated} propose a time-aware spectrum-based fault localization approach that weights statements based on their temporal proximity to failures. In this setting, \glspl{fm} could help in the localization with semantic insights extracted from error messages and execution traces.

\begin{center}
{\color{black}
\small
\renewcommand{\arraystretch}{1.3}
\begin{longtable}{|p{0.06\textwidth}|p{0.23\textwidth}|p{0.27\textwidth}|p{0.12\textwidth}|p{0.20\textwidth}|}
\caption{\revision{Actionable Research Opportunities Corresponding to the Identified Challenges for \gls{cps} Debugging and Repair with Foundation Models}} \label{tab:fm-debugging-repair-opportunities} \\
\hline
\rowcolor{black!70}
\color{white}\textbf{ID.} & \color{white}\textbf{Research Question} & \color{white}\textbf{Methodology} & \color{white}\textbf{Baseline(s)} & \color{white}\textbf{Evaluation Metrics} \\
\hline
\endfirsthead
\caption[]{(continued)} \\
\hline
\rowcolor{black!70}
\color{white}\textbf{ID.} & \color{white}\textbf{Research Question} & \color{white}\textbf{Methodology} & \color{white}\textbf{Baseline(s)} & \color{white}\textbf{Evaluation Metrics} \\
\hline
\endhead
\hline
\endfoot

\rowcolor{white}
$\mathrm{Ch}_{\mathrm{DR1}}$ & How can \glspl{fm} be used for multi-modal temporal fault localization in \gls{cps}? & Fine-tune \gls{fm} pipelines that analyze test logs, sensor data, etc., over a period of time to pinpoint faults. & Baseline \glspl{fm}, Traditional Fault Localization approaches. & Fault localization accuracy, time to locate faults, line coverage. \\
\rowcolor{gray!20}
$\mathrm{Ch}_{\mathrm{DR1}}$ & How can domain knowledge and safety constraints be integrated into \glspl{fm} to improve \gls{cps} fault localization? & Fine-tune \glspl{fm} with safety rules from specifications, regulations, and standards, domain knowledge encoded in ontologies, and historical fault patterns, and enforce these constraints during analysis. & Baseline \gls{fm}. & Improvement in localization accuracy, compliance with safety rules, constraint correctness. \\
\rowcolor{white}
$\mathrm{Ch}_{\mathrm{DR1}}$ & How can \glspl{fm} be combined with spectrum-based fault localization techniques to improve fault localization? & Integrate \glspl{fm} with spectrum-based metrics to prioritize and localize faults, and feed the results into an automated repair implementation. & Spectrum-based approaches alone, baseline \gls{fm}. & Fault detection and localization accuracy, patch success rate, reduction in debugging time. \\
\rowcolor{gray!20}
$\mathrm{Ch}_{\mathrm{DR2}}$ & How can hybrid \gls{fm}-based and search/constraint-based techniques automatically generate patches for \gls{cps} bugs? & Develop pipelines combining \glspl{fm} for code understanding with search/constraint solvers for patch synthesis. & Search/constraint-based techniques, baseline \glspl{fm}. & Patch correctness, coverage of bug types, reduction in manual repair effort, execution time. \\
\rowcolor{white}
$\mathrm{Ch}_{\mathrm{DR2}}$ & How can \gls{fm}-based multi-agent \gls{cps} repair algorithms collaboratively generate and validate patches? & Design multi-agent \glspl{fm} where each agent proposes, tests, and refines patches collaboratively. & Traditional repair algorithms, single \gls{fm}, baseline \gls{fm}. & Patch quality, collaboration efficiency, reduction in repair cycles, fault coverage, hallucination rate. \\
\rowcolor{gray!20}
$\mathrm{Ch}_{\mathrm{DR3}}$ & How can \gls{fm}-based multi-modal surrogate test oracles approximate expensive system-level tests? & Fine-tune \glspl{fm} on historical system-level test results to predict outcomes, with uncertainty estimation. & Baseline \glspl{fm}. & Oracle accuracy, precision and recall, reduction in test execution cost. \\
\rowcolor{white}
$\mathrm{Ch}_{\mathrm{DR3}}$ & How can \glspl{fm} model uncertainty in \gls{cps} validation to reduce test flakiness? & Fine-tune \glspl{fm} to quantify both types of uncertainty and integrate them into testing pipelines. & Baseline \glspl{fm}. & Reduction in flakiness of test results, reliability of test outcomes, uncertainty quantification. \\
\rowcolor{gray!20}
$\mathrm{Ch}_{\mathrm{DR4}}$ & How can temporal-aware fitness functions with \glspl{fm} enable more informative search guidance? & Fine-tune \glspl{fm} on temporal patterns from logs and system behavior to guide repair algorithms. & Baseline search algorithm. & Improvement in early fault detection, convergence speed of search-based repair, and quality of proposed patches. \\
\rowcolor{white}
$\mathrm{Ch}_{\mathrm{DR4}}$ & How can \gls{fm}-based fitness functions go beyond pass/fail to capture continuous behavior and multiple objectives? & Fine-tune \glspl{fm} that evaluate repairs using continuous embeddings representing system behavior and multiple objectives (e.g., safety, performance, reliability). & Traditional binary fitness functions. & Expressiveness of fitness function, repair quality across multiple objectives, reduction in discarded partial solutions, time to repair. \\

\end{longtable}
}
\end{center}

\subsubsection{Fixing a Wider Range of Bugs} Traditional \gls{apr} methods are only able to repair certain types of bugs. For instance, in the context of \glspl{cps}, search-based methods have been proposed to repair misconfigurations~\cite{valle2023automated}, stateflow models~\cite{arrieta2024search} and feature interaction failures~\cite{abdessalem2020automated}. In contrast, \glspl{cps} cover a wide range of bug types~\cite{zampetti2022empirical}, such as, bugs in configuration files. As \glspl{fm} have been trained on massive datasets, we envision their combination with other techniques (e.g., search-based approaches), can significantly help repair a wide range of \glspl{cps} bugs. One promising research direction is domain-specific fine-tuning, where \glspl{fm} are trained on \gls{cps}-relevant codebases, such as Simulink models, PLC code or ROS programs. We conjecture that fine-tuning \glspl{fm} with CPS-specific codebases can help the model learn how control logic is typically structured, enabling it to suggest meaningful patches in these contexts. 

 Another potential research avenue is the development of hybrid repair strategies, i.e., strategies that combine generative capabilities of~\glspl{fm} with search or constraint-based repair techniques. Valle~\cite{valle2025automated} describes an approach in which \glspl{llm} act as mutators within a search algorithm: the search process maintains a population of partial patches and the \gls{llm} is required to generate patches with the help of the information from failure diagnostics. This way, the \gls{fm} is guided by reducing the search space of possible patch solutions. Another research direction we envision is the Multi-Agent Collaboration. For instance, a software repair agent could check code mutations for syntactic correctness and interface validity, while a configuration agent proposes some parameter adjustments. At the same time, an ML auditor could suggest model-centric mitigation such as fine-tuning or calibration. A coordinating agent would mediate between these proposals, resolving conflicts and designing focused validation plans that address the most critical scenarios of the system. For instance, Mapcoder~\cite{islam2024mapcoder}, a multi-agent approach, has shown to be effective at code generation for competitive problem solving.

\subsubsection{Cheaper debugging and repair}
The need for testing \glspl{cps} at the system level is frequently the slowest and most expensive part of \glspl{cps} development process. An individual test can can take minutes to hours~\cite{gladisch2019experience} to execute, require specialized hardware or human supervision, and must often be repeated across many initial conditions to exercise timing-dependent or rare faults~\cite{amini2024evaluating}. This also is exacerbated due to the stochastic nature of such systems. The combination of time-dependant processes and high sensitivity to previous inputs and outputs, could lead the \gls{cps} behave differently across several runs of the same input~\cite{osikowicz2025empirically, amini2024evaluating}. The consequence is that the repair algorithms either cannot search deeply in the search space or consume prohibitive amount of resources.

  \glspl{fm} offer a promising avenue to break this bottleneck by providing fast, informative substitutes for expensive tests and by focusing the execution of real tests where they are most useful. One direction is to develop multi-modal surrogate test oracles~\cite{braga2018machine}. \glspl{fm} trained on historical pairs of short, compact descriptors (e.g., code diffs, scenario parameters) and long-test outcomes can learn to predict the test outcome, safe margins, or even full execution summaries with uncertainty quantification~\cite{malinin2019uncertainty}. Integrating \glspl{fm} in the repair cycle as pre-filter permits the repair loop to discard many low-value candidates cheaply (i.e., the outcome of the surrogate model is that the tests may fail) and reserve the resources for the high-value changes that the surrogate models either ranks the plausible patch as passed or uncertain. Beyond acting as oracle substitutes, \glspl{fm} can also play a central role in handling the test flakiness when evaluating the behavior of \glspl{cps}~\cite{osikowicz2025empirically, amini2024evaluating}. This could be carried out by learning distributions over possible outcomes rather than just predictions. This way, \glspl{fm} can contribute to both efficiency and robustness in navigating the complex, probabilistic behavior space of \glspl{cps}.

\subsubsection{Guiding Repair algorithms}
Search-based repair methods are only as effective as the fitness functions that guide their exploration~\cite{le2012systematic}. Current \gls{apr} approaches~\cite{xia2023automated,yuan2018arja,le2019automated} use the number of failed and passed test cases as a fitness function. While this strategy can be effective in software domains with large, diverse test suites, it is problematic in \glspl{cps} where the number of available test cases is limited and tests often provide little insight into the internal behavior of runs. As a result, the search process may become nearly random, wasting evaluations on uninformative or misleading fitness values.

  \glspl{fm} provide an opportunity to design richer and more informative fitness functions that capture the behavior of \glspl{cps} beyond simple pass/fail outcomes. By learning embeddings of multi-modal execution traces, including telemetry, control signals, timing logs, and environment states, \glspl{fm} can map repairs to continuous behavioral space. This reduces the risk of deceptive signals, where small changes partially satisfy a test but leave underlying faults unaddressed.Moreover, \glspl{fm} can support multi-objective fitness by predicting system-level properties relevant to \glspl{cps}, such as stability margins, latency, or energy consumption. This enables Pareto-based repair strategies~\cite{deb2002fast} that balance competing objectives rather than overfitting to a single binary score. Another advantage is that \glspl{fm} can also incorporate temporal aspects of faults, such as how long a fault has been active or whether early warning signs appear in execution traces. Large models trained on historical runs can learn to anticipate faults before they fully manifest, providing earlier and more informative fitness signals that guide search toward truly corrective patches. Finally, uncertainty-aware \glspl{fm} can actively direct exploration by signaling when predictions are unreliable. This makes it possible to prioritize scare tests for candidate patches that are either high promising or fall into regions of high uncertainty, ensuring that repair resources are allocated effectively.

\section{Foundation Models for \gls{cps} Evolution} \label{sec:CPSevolution}
\revision{This section investigates the role of \glspl{fm} in supporting the evolution of \glspl{cps}. An overview of the necessary background and the state of the art is provided in Section~\ref{subsec:Ev_back}. Subsequently, Section~\ref{subsec:Ev_challenges} identifies key challenges arising from \gls{cps} evolution and corresponding research opportunities, outlining actionable research opportunities and potential directions for addressing these challenges.
}

\revision{
\subsection{Background and State of the Art}\label{subsec:Ev_back}

Recent advances in \glspl{fm} have shown promising potential for supporting software evolution tasks across the software lifecycle. Existing vision and survey papers highlight that current \gls{fm}-based approaches, primarily through \glspl{llm}, can assist with several maintenance and evolution activities, including refactoring, migration, bug fixing, testing, documentation, and change understanding~\cite{marchezan2024model,fan2023large}.
Within model-based software engineering, early research investigates the use of \glspl{fm} for tasks such as model evolution, model generation, consistency checking, and traceability across models and code~\cite{marchezan2024model,di2025use,ferrari2024model,kazai2022model}. These works argue that \glspl{fm} are well suited for evolution tasks that require reasoning over heterogeneous artifacts and global system context rather than isolated code generation.
In the general software engineering domain, \glspl{fm} have been studied for code refactoring, demonstrating their ability to improve code structure and maintainability in legacy systems~\cite{cordeiro2024empirical}. Other works investigate their use for API and library migration, showing support for framework upgrades and platform transitions that are central to software evolution~\cite{almeida2024automatic,cheng2025codemenv}. 
\gls{fm}-based approaches have also been applied to automated program repair, where patches are generated based on failing tests or bug reports, supporting corrective maintenance~\cite{yang2025survey,zubair2025use}. 
Furthermore, empirical studies show that \glspl{fm} can assist in unit test generation and improve test coverage, contributing to regression testing during software evolution~\cite{schafer2023empirical,yang2024evaluation}.
Beyond code-centric tasks, \glspl{fm} have been explored for evolution related artifacts such as commit messages, documentation, change impact analysis, and traceability recovery. These works indicate that \glspl{fm} can support change understanding by summarizing modifications, maintaining documentation consistency, and identifying dependencies across requirements, models, and code artifacts~\cite{lopes2024commit,xue2024automated,zhang2024using,della2024using,viger2024supporting,ali2024establishing,hassine2024llm}. 

Despite these advances, existing FM-based approaches primarily target general-purpose software systems. Their applicability to \glspl{cps} evolution remains limited, as they rarely account for CPS-specific characteristics (e.g., real-time constraints, hardware-software coupling) or support system-level reasoning across architectural layers and cross-module dependencies~\cite{yue2023evolve,al2016software,torngren2018complexity}.
}




\revision{
\subsection{Challenges and Research Opportunities}\label{subsec:Ev_challenges}
This section identifies key challenges associated with the use of \glspl{fm} in supporting the evolution of \glspl{cps}, as summarized in Table~\ref{tab:evolution-challenges}. For each identified challenge, we discuss corresponding research opportunities, including relevant research questions, potential methodologies, and baselines and metrics for evaluating the effectiveness of \gls{fm}-based solutions. An overview of the relationships between challenges and research opportunities is provided in Table~\ref{tab:fm-evolution-opportunities}.
}

\begin{table*}[h]
{\color{black}
\centering
\small
\caption{Challenges in Applying Foundation Models to \gls{cps} Evolution}
\label{tab:evolution-challenges}
\rowcolors{2}{gray!20}{white}
\begin{tabular}{p{0.15\textwidth} p{0.8\textwidth}}
\rowcolor{black!70}
\textcolor{white}{\textbf{Challenge ID}} &
\textcolor{white}{\textbf{Challenge Description}} \\

$\mathrm{Ch}_{\mathrm{E1}}$ &
Handling real-time and data constraints is complicated as CPSs continually evolve and, in some cases, must meet stringent timing requirements. \\

$\mathrm{Ch}_{\mathrm{E2}}$ &
Dealing with complex cross-module dependencies in non-trivial situations, given interactions among heterogeneous components including software, AI modules, hardware, and the physical environment. \\

$\mathrm{Ch}_{\mathrm{E3}}$ &
Ensuring reliability and fault tolerance of evolving CPSs is demanding, as it requires maintaining safe operation while implementing updates. \\

\end{tabular}
}
\end{table*}

\begin{table}[htbp]
{\color{black}
\centering
\footnotesize
\caption{Actionable Research Opportunities in Evolution of \gls{cps} with Foundation Models Corresponding to the Identified Challenges}
\label{tab:fm-evolution-opportunities}
\renewcommand{\arraystretch}{1.3} 
\begin{tabular}{|p{0.05\textwidth}|p{0.17\textwidth}|p{0.30\textwidth}|p{0.15\textwidth}|p{0.20\textwidth}|}
\hline
\rowcolor{black!70}
\color{white}\textbf{ID(s).} & \color{white}\textbf{Research Question} & \color{white}\textbf{Methodology} & \color{white}\textbf{Baseline(s)} & \color{white}\textbf{Evaluation Metrics} \\
\hline
\rowcolor{white}
$\mathrm{Ch}_{\mathrm{E1}}$; $\mathrm{Ch}_{\mathrm{E3}}$ & How can \glspl{fm} support run-time fault detection and self-healing to reduce downtime and manual intervention? & 
Train \glspl{fm} for multimodal runtime anomaly detection using operational logs and integrate them with self-healing controllers to recommend or trigger repair actions. 
& Baseline \glspl{fm}, classical anomaly detection & Mean time to recovery, reduction in system downtime, accuracy of fault detection. \\

\rowcolor{gray!20}
$\mathrm{Ch}_{\mathrm{E1}}$; $\mathrm{Ch}_{\mathrm{E3}}$ & How can \glspl{fm} enable proactive \gls{cps} evolution by predicting issues and suggesting adaptations before failures occur? & Fine-tune \glspl{fm} on historical faults and system behaviors to forecast potential issues and propose preventive adaptations. & Baseline \glspl{fm}, reactive maintenance, threshold-based methods & Prediction accuracy, reduction in unplanned downtime, effectiveness of suggested adaptations. \\

\rowcolor{white}
$\mathrm{Ch}_{\mathrm{E1}}$ & How can \glspl{fm} automatically evolve \gls{cps} digital twins to reflect system changes using multimodal data? & 
Use \glspl{fm} to detect and align heterogeneous system artifacts, including sensor data, logs, and system models, enabling continuous DT synchronization and automated model updates as system behavior evolves.
& Manual, rule-based synchronization & Fidelity of evolved digital twins, accuracy in reflecting system changes, state estimation error (e.g., RMSE, MAE), reduction in update time. \\

\rowcolor{gray!20}
$\mathrm{Ch}_{\mathrm{E2}}$ & How can \glspl{fm} act as autonomous agents to analyze dependencies, predict failures, and safely apply software updates in \gls{cps}? &
Design \gls{fm}-based agents that infer and maintain dependency structures across \gls{cps} modules, predict risky updates, and suggest safe deployment strategies. 
& Non-FM (e.g., model-based and heuristic-based agents) & Accuracy of dependency analysis, prediction of failure-prone updates, reduction in deployment errors. \\

\rowcolor{white}
$\mathrm{Ch}_{\mathrm{E2}}$ & How can \glspl{fm} automatically manage dependencies, update scheduling, and optimize deployment of \gls{cps} software updates? & 
Combine \glspl{fm} with multi-objective optimization techniques to support reasoning over update alternatives, assisting in generating update plans and scheduling deployments that balance safety, timing, and resolve conflicts across dependent modules.
& Baseline \glspl{fm}, manual scheduling & Reduction in update conflicts, deployment efficiency, constraint satisfaction rate. \\

\rowcolor{gray!20}
$\mathrm{Ch}_{\mathrm{E2}}$ & How can \glspl{fm} automate test case evolution for continuous quality assurance in evolving \gls{cps}? & Fine-tune \gls{fm} pipelines to generate, adapt, and prioritize test cases based on system changes and historical test results. & Static regression suites, baseline \glspl{fm} & Test coverage over time, detection of regression faults, fault detection rate, reduction in manual test design effort. \\

\rowcolor{white}
$\mathrm{Ch}_{\mathrm{E3}}$ & How can \glspl{fm} automate long-term \gls{cps} maintenance while ensuring regulatory and standard compliance? & Develop \gls{fm}-based maintenance pipelines integrating compliance checks and adaptation of software components. & Manual audits, rule-based compliance tools & Compliance coverage, reduction in manual maintenance effort, adherence to regulatory standards. \\
\hline
\end{tabular}
}
\end{table}

\subsubsection{Automating \gls{cps} Software Changes and Adjustments}
\revision{Software changes (e.g., due to optimization, feature updates, or new hardware) introduce risks of system instability and compatibility issues, primarily due to hidden dependencies and lack of automated support~\cite{fan2024detect}.} Developers must manually navigate large and complex codebases, ensure compatibility, and prevent unintended behaviors (e.g., new change breaks existing functionality), making the process time-consuming and error-prone. \revision{Similar to GitHub Copilot~\cite{githubcopilot}}, the future of \glspl{fm} in \gls{cps} software evolution extends beyond simply suggesting code changes. Instead of relying on human intervention for execution, \glspl{fm} could be used as autonomous decision-making models capable of analyzing dependencies, predicting potential failures, and applying optimized updates without breaking existing functionality.

\subsubsection{Scalable and Efficient \gls{cps} Software Updates and Deployment}
Managing software updates is challenging, especially in \glspl{cps} that require continuous operation and minimal downtime. Updates must be carefully coordinated to ensure compatibility with existing components, avoid disruptions, and prevent failures. \revision{In complex \glspl{cps}, manually handling dependencies and version control slows the process and increases the risk of errors~\cite{dobaj2023towards}.} To this end, more research is needed to explore to what extent \glspl{fm} can improve software update processes by automating update scheduling, resolving dependencies, and optimizing deployment strategies.

Beyond automation, \glspl{fm} could also support adaptive decision-making, where update strategies are tailored to the operational context (e.g., prioritizing urgent security patches). \revision{Another promising direction is the use of \glspl{fm} to simulate and validate updates before deployment, thus reducing the risks of unexpected failures~\cite{nouri2025simulation}.} 

\subsubsection{Self-healing and Adaptive Mechanisms}
\gls{cps} software failures often require manual intervention, causing downtime and increasing maintenance costs. \revision{Traditional recovery methods, such as system reboots or checkpoint rollbacks, can restore functionality but do not prevent recurring failures or adapt to new issues~\cite{krupitzer2015survey}.} 
In this context, \glspl{fm} can assist by analyzing \gls{cps} behavior, detecting faults early, and providing real-time insights to support adaptive decision-making. 
While \glspl{cps} can integrate predefined recovery strategies, \revision{\glspl{fm} enhance these processes by identifying patterns, diagnosing potential issues, and suggesting corrective actions, reducing reliance on manual intervention~\cite{rauba2024self}.} 

\subsubsection{Automated \gls{cps} Test Evolution and Quality Assurance}
Keeping \glspl{cps} reliable requires testing to evolve alongside new updates and features. 
Both manual and automated \gls{cps} testing methods often fall behind rapid software changes, requiring significant human effort to create and/or update test cases and maintain regression suites. 
\revision{To this end, \glspl{fm} are needed to generate test cases that reflect software changes and maintain realism and relevance~\cite{wu2024reality}.} Moreover, \glspl{fm} can automatically update regression tests, improving test coverage and reliability and enhancing the overall software quality. All these aspects require further research. 

In particular, future work should investigate how \glspl{fm} can be seamlessly integrated into continuous integration and deployment pipelines, enabling automated and on-demand test evolution. Another promising direction is to explore prioritization strategies where \glspl{fm} guides which test cases to execute first, based on criticality or likelihood of failure. \revision{Furthermore, applying \glspl{fm} to capture not only functional but also non-functional aspects (e.g., performance, energy usage) could extend their value for \gls{cps} quality assurance~\cite{dandotiya2025generative}.}

\subsubsection{Proactive Evolution of \glspl{cps}}
Traditionally, the evolution of \glspl{cps} has been reactive: updates and adjustments are introduced in response to failures, changing requirements, or environmental shifts. This leads to costly downtime, delayed adaptations, and vulnerability to unforeseen conditions. \Glspl{fm} open the possibility for proactive evolution, where systems anticipate problems and adapt before they escalate into failures. 
Recent work has already moved toward proactive monitoring and diagnosis in related areas, including predictive anomaly detection~\cite{liang2024foundation}, invariant learning for fault detection~\cite{abshari2024llm}, and \gls{llm}-based fault localization~\cite{yang2024large}. However, these efforts stop at detection and diagnosis. By leveraging runtime data and historical patterns, \Glspl{fm} could support predictive feedback or control loops that suggest timely adjustments and updates to keep systems stable and reliable. To our knowledge, the idea of proactive evolution remains largely unexplored.

\subsubsection{Long-Term Maintenance and Compliance}
\glspl{cps} software must evolve to meet new regulations, security threats, and even hardware changes, but tracking and applying updates is often complex. \glspl{fm} can assist by analyzing compliance requirements, suggesting necessary updates, and automating documentation. Taking the example of a medical dispenser~\cite{sartaj2024modelbased}, an \gls{fm} can review and update software to comply with evolving healthcare standards while adapting to customized settings. Beyond healthcare, \glspl{fm} can be applied in other \gls{cps} domains: AVs can update related modules (e.g., perception) to meet safety certification standards, while industrial robots may need reconfiguration to satisfy both safety and ethical regulations. In all cases, \glspl{fm} reduce manual effort and improve traceability, thereby supporting maintainability, regulatory compliance, and trustworthiness of \gls{cps}. \looseness=-1

\subsubsection{Evolving Digital Twins of \glspl{cps}}
A DT of a \gls{cps} typically consists of a DT model and DT capabilities~\cite{ali2024foundation}. 
Various approaches have been explored in the literature to construct the DT model and its capabilities, which can be developed using the same process or through different ones. 
For instance, the DT model might be created using one type of machine learning (ML) model, while its capability could be developed using model-driven engineering or another ML model~\cite{sartaj2023hita,sartaj2024modelbased}. 
When a \gls{cps} evolves, evolving its DT (including both the model and its capabilities) becomes a significant challenge, particularly when the DT model and capabilities are developed using different methods~\cite{isaku2025digitaltwin,isaku2025oodsar}. 
Recent advances have employed advanced AI techniques, including meta-learning~\cite{sartaj2024medet} and transfer learning~\cite{xu2022uncertainty}, to address these challenges. 
However, these approaches often require substantial amounts of data for training and fine-tuning ML models to ensure that the evolved DT precisely reflects the evolved \gls{cps}. 
This challenge becomes even more critical in domains like ADS and robotics, where multimodal data---visual, sensor, and environmental inputs---is required to be processed. 
\revision{Similar to DTs built using foundation models (Section~\ref{sec:Model4CPS}), these models can play a significant role in evolving DTs~\cite{trantas2024digital}.} 
\revision{One potential direction is to use updated \gls{cps} artifacts and foundation models to generate multimodal data for evolving DTs built using ML models~\cite{chiaro2025generative}.} 
\revision{For the foundation model-based DTs, a potential direction is to develop techniques that use prompt-based fine-tuning to evolve those DTs based on the changed \gls{cps} artifacts, such as requirements, code, or visual data~\cite{ali2024foundation}.} 
\revision{Building on this, another key direction is the exploration of auto-prompting and automated fine-tuning methods~\cite{mei2025survey}} to enable the seamless and automated evolution of foundation model-based DTs, further reducing manual effort.

\revision{
\section{Cross-cutting Challenges when Applying \glspl{fm}}\label{sec:challenges}
In addition to the challenges discussed in each phase, adopting \glspl{fm} in CPS software engineering faces several cross-cutting challenges. Table~\ref{tab:fm-common-challenges-taxonomy} classifies these challenges into six dimensions. These dimensions were selected to provide a holistic view of \gls{fm} adoption barriers, drawing on established frameworks for FM development~\cite{weidinger2021ethical} and covering challenges inherent to \gls{fm} technology, safety and certification, practical deployment constraints, concerned with human and organizational factors, ethical and privacy concerns, and environmental sustainability considerations.
%
For each dimension, we identify key challenges and outline corresponding research opportunities to guide future work. Note that we provide relevant references from which we identified the challenges as well as research opportunities, where applicable.


}

\begin{table}[htbp]
{\color{black}
\centering
\small
\caption{\revision{Cross-cutting Challenges in Applying \glspl{fm} to \gls{cps} Software Engineering}}
\label{tab:fm-common-challenges-taxonomy}
\renewcommand{\arraystretch}{1.3}
\resizebox{\textwidth}{!}{
\begin{tabular}{|p{0.15\textwidth}|p{0.48\textwidth}|p{0.47\textwidth}|}
\cline{1-3}
\rowcolor{black!70}
\color{white}\textbf{Dimension} & \color{white}\textbf{Key challenges} & \color{white}\textbf{Research opportunities} \\
\cline{1-3}
\multirow{8}{*}{Technical} & Hallucination and uncertainties~\cite{ji2023survey,zhang2025siren,farquhar2024detecting,huang2025survey} & Hallucination and uncertainty identification, detection and mitigation\\ \cline{2-3}
& Sensitivity to prompt variations~\cite{sclar2023quantifying,lu2022fantastically} & Robust prompting and prompt optimization~\cite{zhou2024robust,li2023robust,pryzant2023automatic}\\ \cline{2-3}
& Generalization to \gls{cps}-specific domain~\cite{guo2024controlagent,kevian2024capabilities} & Data augmentation~\cite{chai2026text}, knowledge distillation~\cite{xu2024survey,yang2025survey}, domain-specific fine-tuning~\cite{gu2021domain}\\ \cline{2-3}
& Artifacts correctness and transformation~\cite{peng2023towards,tufek2024validating} & Validation of \gls{fm}-generated artifacts~\cite{tufek2024validating}\\ \cline{2-3}
& Real-time inference constraints~\cite{niu2024rtil,zhu2024survey} & Model compression and optimization~\cite{wang2024model,zhu2024survey,du2026survey}\\ \cline{2-3}
& Alignment with human preference~\cite{liu2023trustworthy,shen2023large,wang2023aligning,huang2024trustllm} & Reinforcement learning from human feedback and human-in-the-loop fine-tuning~\cite{bai2022training,wang2023aligning}, preference learning~\cite{muldrew2024active,jiang2025survey}\\ \cline{2-3}
& \gls{fm} customization, collaboration, and reusability~\cite{chen2024large,wang2025comprehensive,huang2024ensemble,ye2024mplug} & Parameter-efficient fine-tuning~\cite{wang2025parameter} and multi-agent collaboration~\cite{qian2024scaling,li2023theory}\\ \cline{2-3}
& Toolchain integration complexity~\cite{10.1145/3663529.3663849,han2024review,zhan2024injecagent} & \gls{fm} supply chain~\cite{10.1145/3708531}\\ \cline{1-3}
%
\multirow{4}{*}{\shortstack[l]{Safety \& \\Certification}} 
& Unverified FM behaviors in safety-critical applications~\cite{gu2025improve,wu2026foundation,cui2024survey,liu2024survey} & Testing and evaluation of \glspl{fm}~\cite{chang2024survey,dobslaw2025challenges}\\ \cline{2-3}
& Incompatibility with existing safety standards~\cite{ong2024ethical,genna2024regulation,10.5555/3716662.3716728} & Extending or developing safety standards and regulatory frameworks\\ \cline{2-3}
& Difficulty of audit certification~\cite{mokander2024auditing,FMOppor} & Explainability and transparency for \gls{fm} certification~\cite{scharowski2023certification,balasubramaniam2022transparency}\\ 
%
\cline{1-3}
\multirow{5}{*}{\shortstack[l]{Economic \& \\Resource}} & High cost in dataset creation~\cite{liu2024datasetslargelanguagemodels,xu2024llm,acquaah2025realistic} & Synthetic data generation~\cite{gholami2023doessyntheticdatamake}\\ \cline{2-3}
& High computational cost of training and fine-tuning~\cite{xu2025resource,hoffmann2022empirical,bao2023tallrec} & Parameter-efficient fine-tuning~\cite{wang2025parameter} and data-efficient training~\cite{sachdeva2024train,luo2025survey}\\ \cline{2-3}
& High inference cost at scale~\cite{samsi2023words,liang2023holisticevaluationlanguagemodels,wang2023cost} & Cost-efficient inference via batch prompting and model compression~\cite{wang2024model,cheng2023batch}\\ \cline{2-3}
& Resource constraints for edge \gls{cps} deployment~\cite{10.1145/3706418,qu2025mobile,zheng2025review} & Model compression, knowledge distillation, and hardware optimization~\cite{wang2024model,10.1145/3719664}\\ \cline{2-3}
& Continuous monitoring and maintenance overhead~\cite{rosen2026portability,baris2025foundation} & \gls{fm} monitoring and lifelong learning strategies~\cite{wang2024assessing,zheng2026lifelong,zheng2025towards}\\
\cline{1-3}
\multirow{4}{*}{\shortstack[l]{Human, \\Organizational \& \\ Social}} & Overreliance on \gls{fm} without critical evaluation~\cite{10.1145/3772318.3790332,10.1145/3449287,jeon2025empowering} & Informed \gls{fm} outputs via transparent explanations and justifications~\cite{kim2025fostering} \\ \cline{2-3}
& Communication gaps in multidisciplinary development teams~\cite{10.1145/3449205} & Explainable \gls{fm} interfaces to bridge communication across disciplines~\cite{boyko2023interdisciplinary,kong2024explainable}\\ \cline{2-3}
& Absence of organizational governance frameworks~\cite{bommasani2024considerations,roberts2024global} & Governance frameworks for \glspl{fm}~\cite{jernite2022data}\\ \cline{2-3}
& Intrinsic bias in FM-generated outputs~\cite{Myers2024,menzner2025bias} & Bias detection and mitigation~\cite{guo2024bias,zhou2024mitigating}\\
\cline{1-3}
\multirow{4}{*}{\shortstack[l]{Ethical \& \\Privacy}} & Ethical biases throughout \gls{fm} lifecycle~\cite{jha2025ethical,abdulhai2024moral,myers2024foundation} & Ethical framework to ensure fairness, transparency, accountability~\cite{head2023large,weidinger2021ethical}\\ \cline{2-3}
& Sensitive \gls{cps} data leakage through \gls{fm} pipeline~\cite{deng2023benchmark,inan2021training,sperli2025exploring} & Differential privacy and data sanitization for \gls{cps} \gls{fm} pipelines~\cite{shi2022just,behnia2022ew,carlini2021extracting}\\ \cline{2-3}
& Proprietary asset exposure through cloud-based \gls{fm} APIs~\cite{carlini2021extracting,balloccu2024leak,spina2025peeking} & On-premise deployment of \glspl{fm}~\cite{zhang2025cloud}\\ \cline{2-3}
& Intellectual property (IP) ownership of \gls{fm}-generated artifacts~\cite{da2025intellectual,makridis2025governing} & Legal and technical frameworks for IP (e.g., EU AI Act)~\cite{act2024eu}\\ 
\cline{1-3}
\multirow{2}{*}{Environmental} & Substantial carbon footprint from \gls{fm} training~\cite{liu2024green,iftikhar2024reducing,jeanquartier2026assessing} & Parameter-efficient fine-tuning~\cite{wang2025parameter}, data-efficient training~\cite{sachdeva2024train,luo2025survey}, model compression~\cite{wang2024model}, and carbon tracking and reporting~\cite{schwartz2020green}\\ \cline{2-3}
& Conflict between \gls{fm} scaling and environmental sustainability~\cite{singh2025survey,bhaskar2024environment} & Efficient hardware and data centers, trade-off management through governance and policy\\
\hline
\end{tabular}
}
}
\end{table}


\revision{
\subsection{Technical Aspects}

\subsubsection{Key Challenges}
A fundamental concern is that \glspl{fm} are prone to generating outputs that may be uncertain, incomplete, or incorrect, often referred to as hallucination~\cite{xu2025hallucinationinevitableinnatelimitation}. In the CPS context, such hallucinations can complicate decision-making and validation by making it difficult to determine whether a generated artifact is reliable or requires further human review. Addressing these challenges is critical for ensuring reliability, safety, and traceability across the CPS software lifecycle. Another concern is \glspl{fm} sensitivity to prompt variations, which can lead to inconsistent results from minor changes in input phrasing~\cite{sclar2023quantifying,lu2022fantastically}. Besides, as \glspl{fm} are primarily trained on generic datasets that do not capture the specialized knowledge, the generalizability of \glspl{fm} to \gls{cps}-specific domains remains limited, which can further result in correctness and transformation issues of \gls{fm}-generated artifacts across \gls{cps} software lifecycle~\cite{peng2023towards,tufek2024validating,husain2024can}. Beyond this, \glspl{cps} often operate in real-time settings, such constraints conflict with the computational demands of large \glspl{fm}~\cite{niu2024rtil,zhu2024survey}, limiting their applicability in time-sensitive \gls{cps} operations. Moreover, human expectations are often expressed in natural language, which is inherently vague and ambiguous, making it challenging for \gls{fm} to align with human engineering preferences~\cite{liu2023trustworthy,shen2023large,wang2023aligning,
huang2024trustllm}
Finally, the effective application of \glspl{fm} requires careful customization, human–\glspl{fm} collaboration, and support for cross-domain reusability, each of which introduces challenges such as scarce domain-specific data and high fine-tuning costs, limited interpretability and effective feedback integration, and difficulties in generalizing models consistently across diverse CPS contexts.

\subsubsection{Research Opportunities}

Several research directions are worth investigating for the above technical challenges. For example, developing uncertainty estimation and hallucination detection methods tailored to \glspl{fm} can improve the reliability and trustworthiness of 
\gls{fm} outputs~\cite{shorinwa2025survey,tonmoy2024comprehensive}. 
Robust prompting strategies~\cite{zhou2024robust,li2023robust} and prompt optimization techniques can be applied to reduce sensitivity to input 
variations~\cite{pryzant2023automatic}. To improve domain generalization, data augmentation, knowledge distillation, and domain-specific fine-tuning on \gls{cps} datasets are promising directions~\cite{chai2026text,xu2024survey,yang2025survey,gu2021domain}. 
Besides, developing validation pipelines and traceability techniques is needed to ensure the correctness and semantic preservation of \gls{fm}-generated artifacts across lifecycle stages~\cite{tufek2024validating}. To address real-time inference constraints, techniques such as model compression and optimization can be applied for more efficient deployment of \glspl{fm} in CPS environments~\cite{wang2024model,zhu2024survey,du2026survey}. In addition, alignment with engineering intent can be achieved through reinforcement learning from human feedback, preference learning, and human-in-the-loop fine-tuning~\cite{bai2022training,wang2023aligning,muldrew2024active,jiang2025survey}. Lastly, to support model customization, collaboration, and reusability across \gls{cps} domains, parameter-efficient fine-tuning methods and multi-agent collaboration frameworks are promising directions~\cite{wang2025parameter,qian2024scaling,li2023theory}.

\subsection{Safety and Certification}

\subsubsection{Key Challenges}
\glspl{cps} are typically safety-critical systems, where failures can lead to severe physical, operational, or economic consequences. The integration of \glspl{fm} into software of such systems raises concerns due to their lack of verifiable behavior in unpredictable operating conditions~\cite{gu2025improve,wu2026foundation,cui2024survey,liu2024survey}. In addition, existing safety standards are not fully compatible with \glspl{fm}~\cite{ong2024ethical,genna2024regulation,10.5555/3716662.3716728} and certification is further complicated by the limited transparency of \glspl{fm}, which hinders auditability, traceability, and accountability in safety assurance processes~\cite{mokander2024auditing,FMOppor}.

\subsubsection{Research Opportunities}
Addressing the above challenges requires new assurance paradigms. One promising direction is developing systematic testing and evaluation frameworks to capture emergent behaviors of \glspl{fm} in safety-critical settings~\cite{chang2024survey,dobslaw2025challenges}.
Another important direction is improving explainability and transparency to support audit and certification processes~\cite{scharowski2023certification,balasubramaniam2022transparency,mokander2024auditing}.
Finally, extending existing safety standards or developing new regulatory frameworks tailored for \glspl{fm} is essential. Such frameworks should account for \gls{fm} evolution by re-certifying and compliance checking across the model lifecycle~\cite{ong2024ethical,genna2024regulation}.

\subsection{Economic and Resource Constraints}

\subsubsection{Key Challenges}
The practical adoption of \glspl{fm} also raises concerns regarding economic and resource constraints. \Glspl{fm} are large models that require large-scale datasets and high computational resources for training, fine-tuning, and inference~\cite{liu2024datasetslargelanguagemodels,xu2024llm,acquaah2025realistic,xu2025resource,hoffmann2022empirical,bao2023tallrec}. For instance, in the data preparation phase, \gls{cps} datasets must capture heterogeneous data modalities and require domain expertise for annotation, which is both expensive and time-consuming. At deployment, \glspl{fm} must perform inference for \gls{cps} monitoring, testing, or code generation tasks, which can be economically prohibitive. Such resource constraints are further amplified when deploying on edge devices, where \glspl{fm} must operate on hardware with limited memory, processing power, and real-time constraints~\cite{10.1145/3706418,qu2025mobile,zheng2025review}. Finally, continuous monitoring and maintenance of deployed \glspl{fm} introduces ongoing operational overhead~\cite{rosen2026portability,baris2025foundation}.

\subsubsection{Research Opportunities}
First, to reduce the cost of creating \gls{cps}-specific datasets~\cite {gholami2023doessyntheticdatamake}, synthetic data generation techniques are promising approaches that generate datasets without requiring costly real-world data collection or manual annotation. Besides, parameter-efficient fine-tuning methods~\cite{wang2025parameter} and data-efficient training strategies~\cite{sachdeva2024train,luo2025survey} can significantly reduce the computational cost of adapting \glspl{fm} to \gls{cps} domains without requiring model retraining. To reduce the inference cost of \gls{fm} deployment at an industrial scale, batch prompting and model compression techniques~\cite{wang2024model,cheng2023batch} are promising directions. For edge deployment, model compression, knowledge distillation, and hardware-aware optimization~\cite{wang2024model,10.1145/3719664} can produce compact \gls{fm} variants suitable for edge device deployment. Finally, automated \gls{fm} monitoring pipelines and lifelong learning strategies~\cite{wang2024assessing,zheng2026lifelong,zheng2025towards} can reduce the manual overhead of maintaining \gls{fm}-based \gls{cps} tools as systems evolve.

\subsection{Human, Organizational, and Social Aspects}

\subsubsection{Key Challenges}
A key risk is overreliance on \gls{fm} outputs without sufficient critical evaluation~\cite{10.1145/3772318.3790332,10.1145/3449287,jeon2025empowering}. Engineers or developers may accept \gls{fm}-generated artifacts (e.g., code, test cases, or design models), without verifying their correctness, increasing the risk of undetected errors propagating into safety-critical \gls{cps} systems.
\gls{fm} adoption also reshapes engineers' collaboration patterns. \gls{cps} development usually involves multidisciplinary teams comprising software, hardware, control, and domain engineers, each with distinct expertise and backgrounds. Communication gaps within such teams are already a significant challenge~\cite{10.1145/3449205}, and the introduction of \gls{fm}-based tools adds further complexity, as team members with limited \gls{fm} literacy may struggle to interpret, validate, or challenge \gls{fm}-generated outputs effectively. 
Moreover, the absence of governance frameworks and polices for \gls{fm} adoption may compromise safe and responsible deployment of \glspl{fm}~\cite{bommasani2024considerations,roberts2024global}, and the intrinsic biases in \gls{fm} may influence engineering decisions, potentially leading to biased system behaviors that affect different user groups unequally in \gls{cps} applications.

\subsubsection{Research Opportunities}
To mitigate the overreliance on \glspl{fm}, informed \gls{fm} outputs, supported by transparent explanations and justifications~\cite{kim2025fostering}, can help engineers critically evaluate generated artifacts rather than accepting them directly. Building on this, explainable \gls{fm} interfaces that provide human-readable justifications can further bridge communication gaps in multidisciplinary \gls{cps} teams~\cite{boyko2023interdisciplinary,kong2024explainable}. At the organizational level, governance frameworks defining clear policies for \gls{fm} use, validation, and accountability in \gls{cps} engineering are equally essential~\cite{jernite2022data}. Finally, bias detection and mitigation techniques for \gls{fm}-assisted \gls{cps} can help identify and reduce intrinsic biases before they affect decision-making~\cite{guo2024bias,zhou2024mitigating}.

\subsection{Ethical and Privacy Aspects}

\subsubsection{Key Challenges}
The use of \glspl{fm} in \gls{cps} software engineering introduces ethical and privacy risks arising from both their development and application~\cite{zhou2024ethical}. Ethical concerns emerge throughout the \gls{fm} lifecycle, including the propagation of biases that may affect fairness, transparency, and accountability in \gls{cps} decision-making~\cite{jha2025ethical,abdulhai2024moral,myers2024foundation}. In addition, sensitive \gls{cps} data used in \gls{fm} pipelines may be exposed during training or inference, leading to potential privacy violations~\cite{deng2023benchmark,inan2021training,sperli2025exploring}. The use of cloud-based \gls{fm} APIs further increases the risk of proprietary asset leakage~\cite{carlini2021extracting,balloccu2024leak,spina2025peeking}. Finally, the deployment of \glspl{fm} raises unresolved questions regarding intellectual property ownership of generated artifacts~\cite{da2025intellectual,makridis2025governing}.


\subsubsection{Research Opportunities}
First, addressing ethical risks requires ethical design frameworks that explicitly promote responsible and transparent \gls{fm} use~\cite{head2023large,weidinger2021ethical}, and the privacy risks motivate research into privacy-preserving learning techniques, such as differential privacy and data sanitization in \gls{fm} training and inference pipelines~\cite{shi2022just,behnia2022ew,carlini2021extracting}. Further, secure deployment strategies such as on-premise or isolated \gls{fm} execution environments are promising directions to avoid proprietary asset exposure~\cite{zhang2025cloud}. Moreover, the development of legal and technical frameworks, including emerging regulatory efforts such as the EU AI Act, is needed to clarify intellectual property ownership~\cite{act2024eu}.


\subsection{Environmental Impacts}

\subsubsection{Key Challenges}
The use of \glspl{fm} introduces significant environmental challenges due to their substantial computational requirements. A major concern is the high carbon footprint associated with \gls{fm} training, which arises from large-scale datasets, extensive optimization procedures, and repeated experimentation~\cite{liu2024green,iftikhar2024reducing,jeanquartier2026assessing}. In addition, there is an inherent conflict between the continued scaling of \glspl{fm} and environmental sustainability, as increasing model size and compute demand lead to higher energy consumption~\cite{singh2025survey,bhaskar2024environment}.

\subsubsection{Research Opportunities}

To address the above challenges, several research directions focus on improving efficiency across the \gls{fm} lifecycle. For instance, parameter-efficient fine-tuning, data-efficient training, and model compression techniques can substantially reduce computational cost while maintaining performance~\cite{wang2025parameter,sachdeva2024train,luo2025survey,wang2024model}. Moreover, carbon-tracking and reporting frameworks provide visibility into the environmental footprint of \gls{fm} development and deployment, enabling better measurement for carbon emissions~\cite{schwartz2020green}. 
At a broader level, the adoption of energy-efficient hardware, optimized data center design, and governance-driven trade-off management between model performance and energy consumption represent complementary directions for reducing the environmental impact of the use of \glspl{fm}. 
}

\section{Related Work}\label{sec:relatedwork}

\subsection{Foundation Models for Software Engineering}

With the rapid advancements of FMs, an increasing number of studies have explored their application across various stages of SE. Literature reviews and surveys are crucial means for providing comprehensive insights into the role of FMs within SE~\cite{SEIndApp2024,10.1109/TSE.2024.3368208,LLMs4SESurvey,LLM4SESurvey2,zheng2025towards,wang2025agents}. Wang et al.~\cite{10.1109/TSE.2024.3368208} conducted a systematic review of 102 studies on the use of LLM in software testing. Their work highlights test case preparation and program repair as the most prominent tasks where LLMs have been applied, while also analyzing the key challenges and opportunities in applying LLMs in software testing. Hou et al.~\cite{LLMs4SESurvey} conducted a systematic literature review of 395 studies published between 2017 and early 2024, providing a broader perspective on LLM applications across SE tasks, offering a broader perspective of LLMs across SE tasks, methods, and evaluation strategies. Fan et al.~\cite{LLM4SESurvey2} surveyed the application of LLMs in SE, highlighting their potential across tasks ranging from coding and requirements to documentation and analytics, while also discussing the significant technical challenges posed by LLM properties such as hallucinations. Zheng et al.~\cite{zheng2025towards} systematically reviewed 123 studies from 2022 onward on the integration of LLMs with SE. Their work highlights research focuses, identifies gaps in LLM for SE, and synthesizes evaluations of LLM performance, revealing both capabilities and limitations while providing directions for future research and optimization. 

While the above studies provide comprehensive perspectives of LLM applications in SE, they narrow the scope to a single type of FM, i.e., LLM. As FMs continue to evolve, encompassing multimodal, task-specific, and hybrid models, there is a growing need to examine their broader applications across SE. One relevant study by Li et al.~\cite{SEIndApp2024} examined FM4SE and SE4FM practices in industry from a practitioner's perspective, highlighting prominent SE tasks such as code generation, deployment, and API recommendation. Another relevant work from Shi et al.~\cite{shi2025efficient} presents a roadmap for LLMs in SE, emphasizing the need for efficient and environmentally sustainable solutions. Different from the above work, this paper presents a vision on FM for SE in the domain of CPS, exploring how advanced FMs can support complex, potentially safety-critical, CPS tasks by facilitating the CPS software development cycles, enhancing automation, and addressing domain-specific challenges that go beyond conventional SE applications.

\subsection{Foundation Models for Cyber-physical Systems}

The emergence of FMs has significantly advanced the development of CPSs, providing powerful solutions for ensuring safe and reliable CPS design and operation~\cite{xu2024llm}. This has been demonstrated in a wide range of CPSs, including autonomous vehicles~\cite{Cui_2024_WACV,yang2023llm4drive,li2024large}, robotics~\cite{wang2024large,zeng2023large,kim2024survey,firoozi2025foundation}, and industrial CPSs~\cite{ren2025industrial,song2023pre}. Cui et al.~\cite{Cui_2024_WACV} investigated the use of multimodal FMs in autonomous driving, reviewing tools, datasets, and benchmarks, and identifying key challenges and future directions. Similarly, Yang et al.~\cite{yang2023llm4drive} systematically review the research about FMs for autonomous driving, including the current status of technological advancements, related challenges, and future research directions. Li et al.~\cite{li2024large} provided a comprehensive survey on using LLMs for autonomous driving, examining their applications in both modular pipelines and end-to-end systems. The study highlights LLMs’ potential to enable knowledge-based, human-like autonomous driving, while also addressing key challenges such as real-time inference, safety assurance, and deployment costs. Regarding FMs in robotics, Zeng et al.~\cite{zeng2023large} reviewed the applications of LLMs in robotics, highlighting their impact on control, perception, decision-making, and path planning, while also discussing challenges and future directions for achieving embodied intelligence. Wang et al.~\cite{wang2024large} reviewed the use of FMs in robotics and proposed a GPT-4V-based framework that enhances embodied task planning by combining language instructions with visual perception. Kim et al.~\cite{kim2024survey} surveyed recent advancements of LLMs in robotics, focusing on their applications in communication, perception, planning, and control, and provided practical guidelines and examples for integrating LLMs into robotic systems.

Different from the above studies, this paper focuses specifically on the SE aspects of CPSs, identifying the challenges of applying FMs across the software lifecycles of CPS software, including development, assurance, and maintenance of CPS software, and exploring how FMs can be leveraged to enhance reliability, automation, and efficiency in the CPS software lifecycles. This area has received limited attention in current research.

\revision{
\section{Threats to Validity}\label{sec:conclusions}
\textbf{Construct Validity.} We did not follow a systematic literature review process, and as a result, some challenges and, consequently, some opportunities may not have been captured. To mitigate this, the identification of challenges and opportunities was grounded in an extensive reading of the relevant literature and supplemented by structured group discussions following the Delphi method~\cite{linstone1975delphi} and the nominal group technique~\cite{harvey2012nominal}. In addition, discussions at the SE2030 Workshop (co-located with FSE 2025) provided external input from researchers and practitioners across diverse disciplines, helping to broaden the coverage and balance of the identified challenges. Furthermore, we emphasize that, as a research roadmap, its primary aim is to present representative challenges and opportunities, provide guidance to researchers, and serve as a foundation for future empirical work. Missing some items does not reduce its value, as the roadmap can be iteratively refined as new studies emerge.
%
%
%
\textbf{Conclusion Validity.} As a visionary paper, this work does not include empirical contributions, and the proposed research opportunities have not yet been validated through experiments or case studies. Consequently, their practical impact may differ from what is anticipated. To mitigate this threat and provide structured guidelines for future empirical validation, each research opportunity is accompanied by concrete research questions, suggested methodologies, baseline approaches, and evaluation metrics, ensuring that the roadmap serves not only as a visionary guide but also as an actionable starting point for subsequent empirical investigations.
\textbf{External Validity.} This roadmap is grounded in the current landscape of \glspl{fm}, covering established models such as \glspl{llm}, \glspl{vlm}, and multimodal \glspl{fm}. However, as \gls{fm} evolves, new model paradigms may emerge that are not covered by the current roadmap. For instance, world models~\cite{lecun2022path} represent emerging directions that could open new research opportunities beyond those identified here. As new \gls{fm} paradigms and empirical evidence emerge, the roadmap can be iteratively extended to reflect these developments.
}

\section{Conclusion and Future Work}\label{sec:conclusions}
\revision{This paper presents a forward-looking research roadmap for integrating \glspl{fm} into the software engineering of \glspl{cps}.} It discusses research challenges and corresponding actionable opportunities related to using \glspl{fm} \revision{across six \gls{cps} software engineering phases, including typically phases such as requirement engineering, design and modeling, and testing.} \revision{Each research opportunity presents concrete research questions, suggested methodologies, baseline approaches, and evaluation metrics.} Additionally, \revision{we identify common cross-cutting challenges when adopting \glspl{fm} across six dimensions: technical, safety and certification, economic and resource, human and organizational, ethical and privacy, and environmental, together with corresponding research opportunities for each dimension. This roadmap aims to provide a structured and actionable guide for researchers and practitioners working at the intersection of \glspl{fm} and \gls{cps} software engineering.} In the future, we will expand on these research opportunities in detail for each phase and map existing software engineering problems to various types of \glspl{fm} to present a more comprehensive research roadmap.

\section{Acknowledgements}
This work is supported by the RoboSapiens project funded by the European Commission's Horizon Europe programme under grant agreement number 101133807 and the Co-evolver project (No. 286898/F20) funded by the Research Council of Norway.
Aitor Arrieta and Pablo Valle are part of the Systems and Software Engineering group of Mondragon Unibertsitatea (IT1519-22), supported by the Department of Education, Universities and Research of the Basque Country. Pablo Valle is supported by the Pre-doctoral Program for the Formation of Non-Doctoral Research Staff of the Education Department of the Basque Government (Grant n. PRE\_2024\_1\_0014). Aitor Arrieta is supported by the Spanish Ministry of Science, Innovation and Universities (project PID2023-152979OA-I00), funded by MCIU /AEI /10.13039/501100011033 / FEDER, UE.

\bibliographystyle{plainnat}
\bibliography{references}

\end{document}